\documentclass[showpacs,preprintnumbers,amsmath,amssymb]{revtex4}
\usepackage{epsfig,epstopdf}
\usepackage{dcolumn}% Align table columns on decimal point
\usepackage{bm}% bold math
\usepackage[]{units}
\usepackage[T1]{fontenc}

\usepackage[usenames]{color}
\definecolor{Pierre}{rgb}{0.5,0.,0.5}

\newcommand{\angstrom}{\textup{\AA}}

\newcommand{\remove}[1]{}

\def\be{\begin{equation}}
\def\ee{\end{equation}}

\newcommand{\beq}{\begin{equation}}
\newcommand{\eeq}{\end{equation}}
\newcommand{\beqa}{\begin{eqnarray}}
\newcommand{\eeqa}{\end{eqnarray}}

\newcommand{\bea}{\begin{array}}
\newcommand{\ea}{\end{array}}

\begin{document}

\title{Galaxy cluster constraints on the coupling to photons of low-mass scalars}

\author{Philippe Brax}
\affiliation{Institut de physique théorique, Université Paris Saclay, CEA, CNRS, F-91191 Gif-sur-Yvette
}
\author{Pierre Brun}
\affiliation{CEA, Irfu, Centre de Saclay, F-91191 Gif-sur-Yvette -- France}
\author{Denis Wouters}
\affiliation{CEA, Irfu, Centre de Saclay, F-91191 Gif-sur-Yvette -- France}
\vspace{.2 cm}

\date{\today}
\vspace{.2 cm}

\begin{abstract}

We consider a broad class of interactions between radiation and a light scalar field, including both conformal and disformal couplings. Such a scalar field potentially acts on cosmological scales as dark energy and could also appear in modified gravity theories.
We study the consequences of these couplings on the mixing between the scalar field and photons in galaxy clusters in the presence of a magnetic field. In particular we focus on the resulting turbulence-induced irregularities in the X-ray and UV bands. We find new bounds on the
photon-to-scalar couplings, both conformal and disformal,
which complement laboratory experiments and other astrophysical constraints.

\end{abstract}

 \vskip2pc

\maketitle

\section{Introduction}
\label{Introduction}
Light scalar fields may play a crucial role in the recent history of the Universe. They could be the seeds for the acceleration of the expansion of the Universe and play the role of dark energy at late time \cite{Copeland:2006wr}. They could
modify the properties of gravity on very large scales \cite{Clifton:2011jh}. Their observable properties depend crucially on the type of couplings they have with matter fields and radiation.
Conformal couplings involve a Yukawa-like interaction between scalars and fermions.
Their study is therefore very motivated from a fundamental perspective.
They are responsible for the modification of gravity by scalar fields and explicitly imply the
changes of matter trajectories which are tightly constrained by solar system experiments \cite{Will:2014xja}. On large cosmological scales, the effects of a modification of gravity by conformal couplings could be seen in the growth of structure and will be actively sought for by the Euclid mission \cite{Amendola:2012ys}. These conformal  couplings do not lead to a direct interaction of scalars to photons at tree level as the photon Lagrangian is conformally invariant classically. Such a coupling is only induced at the one loop level when
virtual fermions mediate the interactions between two photons and one scalar \cite{Brax:2010uq}. We will include such a conformal coupling in our analysis. Another type of coupling is particularly important and will be one of the original aspect of this article: disformal couplings \cite{Bekenstein:1992pj,Kaloper:2003yf}. They involve the coupling of scalars to the full energy momentum tensor of matter and radiation. They could arise if gravity becomes massive and  the scalar field arises as the  zero helicity part  of a massive graviton \cite{deRham:2012az}. These couplings involve  two derivatives of the scalar field and can therefore only play a role in dynamical situations. They have no effect on static tests of gravity involving a modification of static forces for instance and are therefore not testable in the solar system \cite{Brax:2012hm}. On the other hand, such a coupling to photons has several consequences. They are very tightly constrained by astrophysics \cite{Brax:2014vva}, collider experiments and cosmology \cite{Brax:2013nsa, vandeBruck:2015ida}. Astrophysically, the coupling to baryons would increase the burning rate of stars and supernovae. Similarly, the primakoff process in stars induced by the disformal coupling to photons implies that a photon in the electric field of a nucleus would release two scalars. This effect allows one to obtain constraints on the suppression scale of the disformal coupling to photons \cite{Brax:2014vva}.
In collider experiments, two quarks would transform into two scalars which would appear as missing energy. Both LHC experiments ATLAS and CMS were able to constraint the coupling scale too \cite{Cembranos:2013qja}.
The disformal coupling plays also a role on large scales  where it would induce  a change of the speed of light which would distort the cosmic microwave background spectrum \cite{Brax:2013nsa}.
In the laboratory, in the presence of a magnetic field, the photons of a laser beam would see their polarisation rotate and become elliptic in a way akin to the well-known effects for Axion-Like-Particles (ALP). In this case, the disformal coupling together with the direct coupling to photons lead to a non-zero transition probability between an initial photon state and a scalar\cite{Brax:2012ie}. The disformal coupling alters the usual transition probability for ALP's. In this article, we will concentrate on the effects of both the conformal and the disformal couplings in astrophysical situations where the external magnetic field appears inside large galaxy clusters for instance.

In galaxy clusters, the magnetic fields are turbulent, as revealed by radio observations. The idea here is to observe a source of radiation (hereafter X-ray photons and UV light) that lies at the center of a galaxy cluster. In the regime where the photons and the scalar field mix at the source, the content of the radiation along the line of sight is actually a mix of photons and scalar field excitations. Because of the turbulent nature of the magnetic field and the energy-dependent mixing, strong irregularities can be introduced in the photon energy spectra, in a way that would resemble a set of random absorption lines. The exact shape of the induced spectral shape is unpredictable, but the statistical properties of the induced noise can be inferred~\cite{2012PhRvD..86d3005W}. This  is used to constraint parameters of the scalar field model in a similar way to  the ALP parameter space (mass and coupling to photons) which has been analysed using gamma-ray sources~\cite{2013PhRvD..88j2003A}, and X-ray sources\cite{2013ApJ...772...44W}.

In this paper, we consider the electrodynamics of disformally coupled scalars in section II. In section III, we apply it to the propagation through turbulent media. In section IV, we deduce constraints on the parameter space
of scalars coupled to photons from X-ray observations, and similarly in section V with UV probes. In section VI, we obtain constraints on the parameter space of the scalar models. We conclude in section VII.

\section{Disformally coupled scalar fields}
\subsection{The scalar models}

We consider the coupling of a light scalar field to matter governed by the action
\be
S=\int d^4x \sqrt{-g}\left(\frac{R}{16\pi G_N} -\frac{1}{2} (\partial \phi)^2 -V(\phi)\right) + S_m(\psi_i, g_{\mu\nu})+ S_\gamma(A_\mu, \tilde g_{\mu\nu})\;,
\label{eq:action}
\ee
where $G_N$ is Newton's constant, which is related to the  reduced Planck scale as $m_{\rm Pl}^2= (8\pi G_N)^{-1}$. The first term in the action is the Einstein-Hilbert action describing General Relativity (GR). The scalar field is canonically normalised with an interaction potential $V(\phi)$ which we do not specify here. In our study we will take
the potential term to be a simple mass term $V(\phi)=\frac{m^2}{2} \phi^2$. This simplified setting could  serve as a template for massive gravity models where the scalar polarisation of the massive graviton becomes a low mass scalar field. More complex cases could be considered in models of screened modified gravity where the potential has a minimum $\bar \phi$ which could depend on the environment, i.e. it would depend on the matter density. Expanding around this minimum $\phi=\bar\phi +\delta \phi$, the potential would reduce, to leading order, to
the case of a massive particle corresponding to the field $\delta \phi$. In the physical situations that we will consider where photons traverse large clusters in the Universe, we will assume that the scalar field in this environment reduces to a massive scalar with a quadratic potential. The matter action $S_m$ describes the matter  part of the standard model of particles physics.
The coupling to photons in $S_\gamma$  directly involves the scalar field. Indeed, this coupling between the scalar and photons is dictated by the disformal metric
\be
\tilde g_{\mu\nu}= g_{\mu\nu} +\frac{2}{M^4} \partial_\mu\phi \partial_\nu \phi\;,
\label{eq:tildemetric}
\ee which will constrained using galaxy cluster observations in the following sections. We have not considered a disformal coupling to matter particles in  $S_m$  as it must be heavily suppressed to comply with the ATLAS constraint. Indeed it is shown in \cite{Brax:2014vva} that the dimensionful disformal coupling to baryons is constrained at the level of $M_b\gtrsim 490 \ {\rm GeV}$. As a result, in this paper, we consider that at the astrophysical energies (X-rays and UV) considered here the disformal coupling to matter is negligible and will be set to zero. Notice that the coupling to different matter species together with the one to photons does not have to be universal. Moreover, the astrophysical and collider constraints obtained in \cite{Brax:2014vva} indicate that disformal couplings have very different effects in different contexts, from laboratory experiments to the burning of stars, and that they are better analysed as separate constants whose values should be inferred from either experiments or observations. In this paper, we focus on the disformal coupling to photons.

The metric $\tilde{g}_{\mu\nu}$ is the Jordan frame metric of photons with respect to which photons are conserved
$
 \tilde D_\mu \tilde T_\gamma^{\mu\nu}=0\;,
$
where the Jordan frame energy momentum tensor is
$
\tilde T^{\mu\nu}= \frac{2}{\sqrt{-\tilde g}} \frac{\delta S_m}{\delta \tilde g_{\mu\nu}}\;.
$
On the other hand, the metric $g_{\mu\nu}$ defines the Einstein frame and in this frame  energy-momentum is not conserved. The Klein-Gordon equation is modified by the presence of the energy momentum of photons
\be
\Box\phi -\frac{2}{M^4}D_\mu (\tilde T_\gamma^{\mu\nu} \partial_\nu \phi) =\frac{\partial V}{\partial \phi}\;.
\label{eq:scalareom}
\ee
The non-conservation of photons, which results from the coupling of the scalar to photons and therefore the exchange of energy-momentum between the scalar and radiation,  appears in the second term of this equation.
In what follows we will restrict ourselves to the leading order effects of the disformal coupling between the scalar field and photons.  Therefore we calculate only to leading order in $1/M^4$, and to this order we have
$\tilde T^{\mu\nu}=T^{\mu\nu}$ where $T^{\mu\nu}$ is the energy-momentum of radiation in the Einstein frame. Higher order terms in $1/M^4$ are present but they would lead to correction terms in $1/M^8$ in the Klein-Gordon equation that we shall neglect. For radiation, this implies that we will use the energy momentum tensor
\be
T_{(\gamma)}^{\mu\nu}= F^{\mu\alpha}{F^\nu_\alpha} -\frac{g^{\mu\nu}}{4} F^2.
\ee
The Klein-Gordon equation reduces to
\be
\Box\phi -\frac{2}{M^4}D_\mu ( T^{\mu\nu} \partial_\nu \phi) =\frac{\partial V}{\partial \phi}\;,
\ee
where we will explicitly take $V(\phi)=\frac{1}{2}m^2\phi^2$.
At this order, the action can be expanded as
\be
S=\int d^4x \sqrt{-g}\left(\frac{R}{16\pi G_N} -\frac{1}{2} (\partial \phi)^2 -V(\phi)+ \frac{1}{M^4} \partial_\mu\phi\partial_\nu\phi T^{\mu\nu}_\gamma \right) + S_m(\psi_i, g_{\mu\nu})\;.\label{eq:GravityFieldAction}
\ee

These models have links with varying speed of light theories which can be found in \cite{Brax:2013nsa}.
The coupling scale $M$ is constant and unknown and should be fixed by observations as we will see.
In certain models where the screening effects appear due to non-linearities in the kinetic terms, the action for the canonically normalised fluctuations $\delta \phi$ around the background value $\bar\phi$ is such that the coupling scale $M$ becomes environment dependent. One may therefore consider that $M$ may depend on the local density. For this reason, we shall quote our new bounds by recalling that they are obtained from galaxy clusters which are much less dense than stars. As a consequence, the star burning constraints may be relaxed, for instance,  in models of the K-mouflage type where stars are screened whereas clusters are not, implying that the suppression scale in stars and in clusters can be very different. A thorough study of the density dependence of the scale $M$ is left for future work. Here we only consider the specific environment provided by galaxy clusters from which sound bounds can be extracted.

In the following, we will also specialise to  the case where  the new scale $M=\sqrt{mm_{\rm Pl}}$. This is motivated by the decoupling limit of the new theories of massive gravity where the scalar field appears as the scalar polarisation of a massive graviton of mass $m$. In this case the
disformal coupling is parameterised by the scale $M$ related to the mass of the graviton $m$. The burning of stars such as the sun can be increased by the disformal coupling to photons as it can lead to a new Primakoff effect whereby scalar escape stars and affect the life-time of stars. This leads to the astrophysical bound $M\gtrsim 346$ MeV, valid in dense environment around $100\ {\rm g/cm^3}$. As already noted, this bound may be relaxed in models where the disformal coupling scale is density dependent. If this is not the case and when $M^2= mm_{\rm Pl}$, this implies that one cannot probe masses lower than $10^{-19}$ GeV which corresponds to scales of order less than 1 km. We will find that the constraints obtained from galaxy clusters where the density is way lower, i.e. $10^{-24}\ {\rm g/cm^3}$, are looser and masses as low as $10^{-31}$ GeV  can be allowed.

\subsection{Electrodynamics with a disformal coupling}

Let us  generalise the previous setting  by introducing a field dependent coupling constant, controlled by a new unknown scale $\Lambda$, so that the kinetic term for photons contains both the disformal coupling and a direct coupling
\be
S_{\rm rad}\supset -\int d^4 x \sqrt{-\tilde g} \frac{1}{4}\left(1+\frac{4\phi}{\Lambda}\right) F^2\;, \label{eq:EDFieldAction}
\ee
where contractions are made with the Jordan frame metric.
The fine structure constant becomes field dependent
$
\alpha (\phi)= \frac{\bar \alpha}{1+\frac{4\phi}{\Lambda}}
$
where $\bar \alpha$ is its value in the absence of the direct coupling. The purely disformal case is obtained by taking $\Lambda\to \infty$.  When the scalar field has a time dependence at the background level $\bar\phi (t)$, this leads to a time variation of the fine structure constant $\alpha(\bar \phi)$ which is constrained at the $10^{-6}$ level at small redshift $z\lesssim 3$ \cite{Evans:2014yva}.
In the following we will consider that the background evolution $\bar\phi$ of the scalar field $\phi$ is such that this constraint is satisfied. Denoting by $\bar \phi_0$ the value of the background field now, we identify the fine structure now as $\alpha=\alpha (\phi_0)$. The variation of the fine structure constant is given by
\be
\frac{\Delta \alpha}{\alpha}= \frac{4\vert \bar\phi-\bar\phi_0\vert}{\Lambda +4 \bar \phi}
\ee
which is small provided that $\vert \bar\phi-\bar\phi_0 \vert \ll \Lambda$ in the recent past of the Universe. In the following, we will use the simplifying assumption that $\bar\phi\ll\Lambda$. Our results can be easily extended to the case when this is not the case anymore.

As mentioned before, the disformal coupling leads to a coupling between the scalar field and photons which becomes, to leading order in $1/M^4$,
\be
{\cal L}= \sqrt{-g}\left( -\frac{1}{4} F^2 -\frac{\phi}{\Lambda} F^2 + \frac{1}{M^4} \partial_\mu\phi\partial_\nu\phi T_{(\gamma)}^{\mu\nu}\right)\;.
\label{eq:photlag}
\ee
This is the action that we shall use to study the mixing between the scalar and photons.
The equation of motion for the photon  gives the generalised form of Maxwell's equation\cite{Brax:2012ie}
 \be
 \partial_\alpha \left[ \left(1+ \frac{4\phi}{\Lambda} +\frac{1}{M^4} (\partial\phi)^2\right)F^{\alpha \beta}\right] -\frac{2}{M^4} \partial_\alpha\left[\partial^\mu\phi\left(\partial^\alpha \phi F_\mu^\beta -\partial^\beta\phi  F_\mu^\alpha\right)\right]=0\;.
 \ee
where the leading term in $1/M^4$ has been kept.
 In the following we consider that  there is a background magnetic field $B_i$  and that we decompose the vector field $A_\mu$ into
 \be
 A_i=   \epsilon_{ijk} B_j x_k + a_i
 \ee
 and $A_0=0$. We work in the Lorentz gauge $\partial_\mu A^\mu=0$ which implies that, in Fourier modes, the vector field $a_i$ is transverse $k_i a^i=0$ with respect to the wave vector $k_i$. Similarly we assume that
 the scalar field takes a background value $\bar \phi$  that we shift
\be
 \phi=\bar \phi +\delta \phi
 \ee
where the fluctuations $\delta \phi$ are massive with a mass $m$. We also assume that the time variation of the background field $\bar\phi$ is negligible in the astrophysical situations we consider, e.g. the propagation of photons through a galaxy cluster.
Defining the dimensionless constants
 \be
 a^2=-\frac{4\bar \phi}{\Lambda},\ \ b=\frac{B^2}{M^4}
 \ee
we find that Maxwell's equation becomes
 \be
 (1-a^2) (E^2 -k^2) a_i -\frac{4i}{\Lambda} \epsilon_{ijk} B_j k_k \delta \phi=0
 \ee
 where $k^2= k_ik^i$ and $E$ is the energy  of the photon represented by the  wave $a_i\sim a_i^{(0)} e^{i(E t - k.x)}$.
 The Klein-Gordon equation becomes
 \be
 \left [(1+b^2) (E^2 -k^2) -m^2 +2b^2  \left ( k^2 - k_i k_j \frac{B^iB^j}{B^2}\right ) \right ]\delta \phi+ \frac{4i}{\Lambda} \epsilon_{ijk} k_i B_j a_k=0
 \ee
 where all the contracted indices involve the Kronecker $\delta_{ij}$.
It is convenient to specialise the propagation along the $z$ axis and normalise
 \be
 \delta \phi \to  \sqrt{1+b^2}\ \delta \phi,\ a_i \to \sqrt{1+a^2}\ a_i.
 \ee
 The mixing between the photons and the scalar field is such that the dispersion relation is still to leading order $E\sim k$, and upon using $k= i\partial_z$ we have
 \be
 (E -i\partial_z)V+ {\cal M} V=0
 \label{eq:equa_motion}
 \ee
 where we have defined the vector
 \be
V=
\left( \begin{array} {c}
a_x\\
a_y\\
i\delta\phi\\
\end{array}\right )
\ee
 and the mixing matrix
\be
{\cal M}=
\left(
\begin{array}{ccc}
0&0& -\frac{2B_y}{\Lambda\sqrt{(1+b^2)(1-a^2)}}\\
0&0&\frac{2B_x}{\Lambda\sqrt{(1+b^2)(1-a^2)}}\\
-\frac{2B_y}{\Lambda\sqrt{(1+b^2)(1-a^2)}}&\frac{2B_x}{\Lambda\sqrt{(1+b^2)(1-a^2)}}&-\frac{m^2}{2E(1+b^2)} +\frac{b^2 E}{1+b^2}(1-\frac{B_z^2}{B^2})\\
\end{array}\right)
\ee
In a plasma the mixing matrix is modified and becomes
\be
{\cal M}=
\left(
\begin{array}{ccc}
-\frac{\omega^2_{\rm pl}}{2E}&0& -\frac{2B_y}{\Lambda\sqrt{(1+b^2)(1-a^2)}}\\
0&-\frac{\omega^2_{\rm pl}}{2E}&\frac{2B_x}{\Lambda\sqrt{(1+b^2)(1-a^2)}}\\
-\frac{2B_y}{\Lambda\sqrt{(1+b^2)(1-a^2)}}&\frac{2B_x}{\Lambda\sqrt{(1+b^2)(1-a^2)}}&-\frac{m^2}{2E(1+b^2)} +\frac{b^2 E}{1+b^2}(1-\frac{B_z^2}{B^2})\\
\end{array}\right)
\label{eq:mixing_matrix}
\ee
where $\omega^2_{pl}= \frac{4\pi \alpha n_e}{m_e}$ is the plasma frequency. Notice that the plasma frequency is not field dependent as we have assumed that $\bar\phi/\Lambda \ll 1$.
This allows one to deduce the
the probability of conversion from a photon to the scalar in a uniform magnetic field after a distance $s$. It  can be exactly calculated and reads
\begin{equation}
P_{\gamma\rightarrow a}(s) = \frac{1}{1+\frac{(\Delta_{\rm pl}-\Delta_{\rm a})^2}{4\Delta_{\rm B}^2}}\sin^2 \frac{2\pi s}{\lambda}
\label{eq:proba}
\end{equation}
with
\begin{eqnarray}
\Delta_{\rm pl} & = & \frac{\omega_{\rm pl}^2}{2E} \\
\Delta_{\rm a} & = & \frac{m^2}{2E(1+b^2)} + \frac{b^2E(1-\frac{B_z^2}{B^2})}{1+b^2} \\
\Delta_{\rm B} & = & \frac{2B}{\Lambda\sqrt{1+b^2}} \\
\lambda & = & \frac{4\pi}{\sqrt{(\Delta_{\rm pl}-\Delta_{\rm a})^2+4\Delta_B^2}}
\end{eqnarray}
When $b\rightarrow 0$, the ALP case is retrieved. In the following we will see that the presence of the new coupling $b$ introduces drastic differences in the energy dependence of the transition probability.

\subsection{Energy dependence}
\label{sec:critical}

\begin{figure}[h]
  \centering	
  \includegraphics[width=0.3\columnwidth]{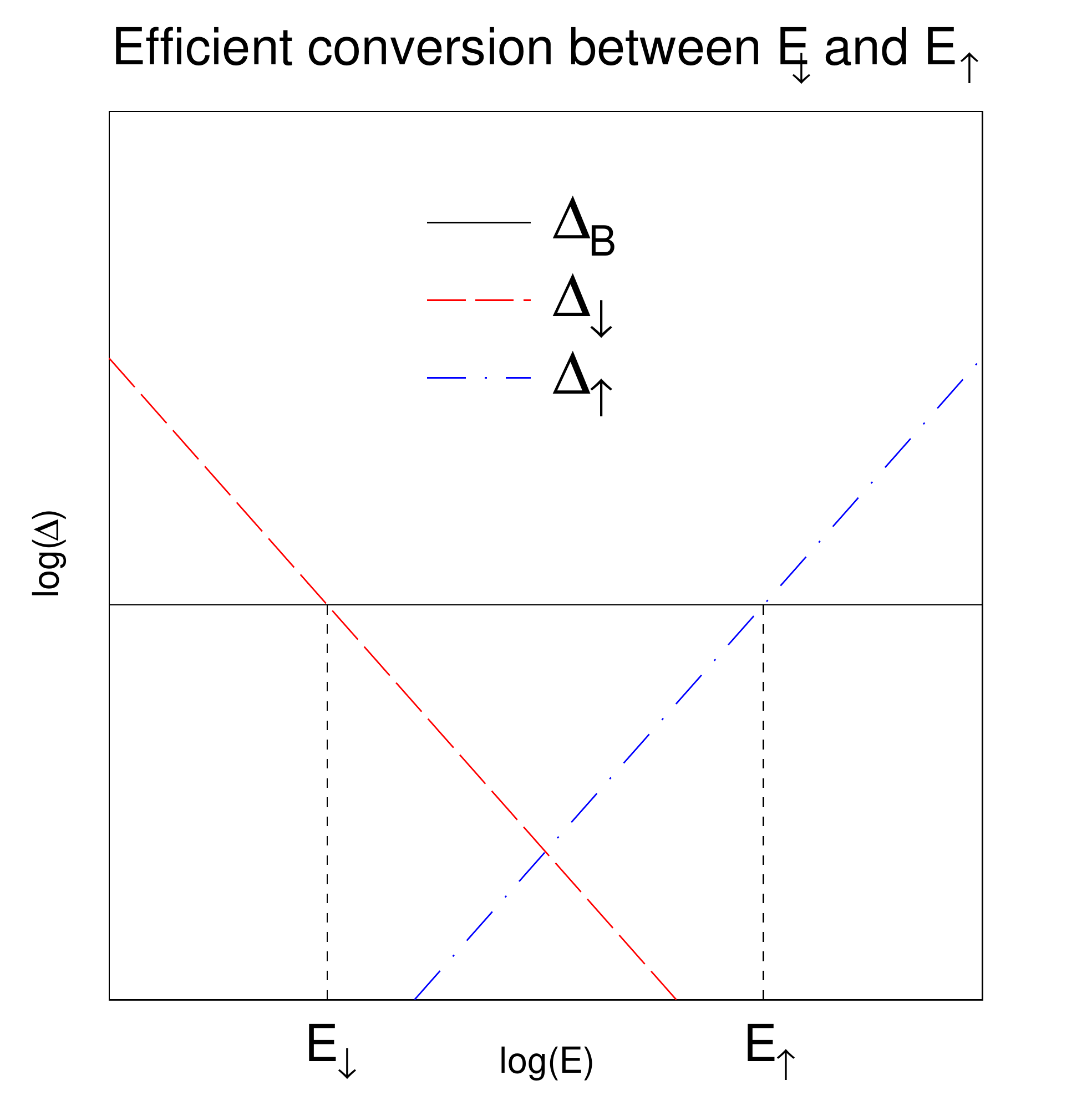}	
 \includegraphics[width=0.3\columnwidth]{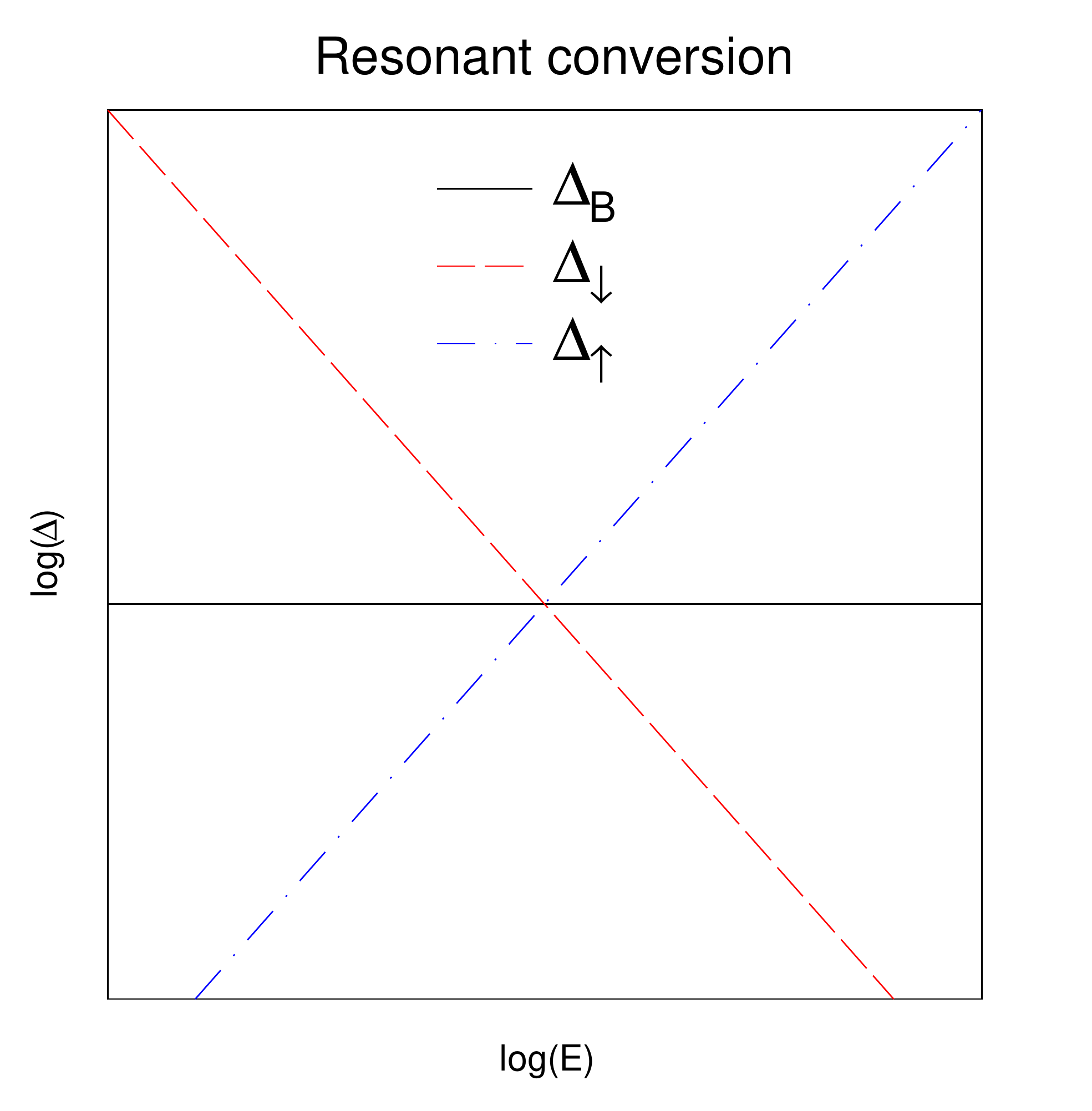}
  \includegraphics[width=0.3\columnwidth]{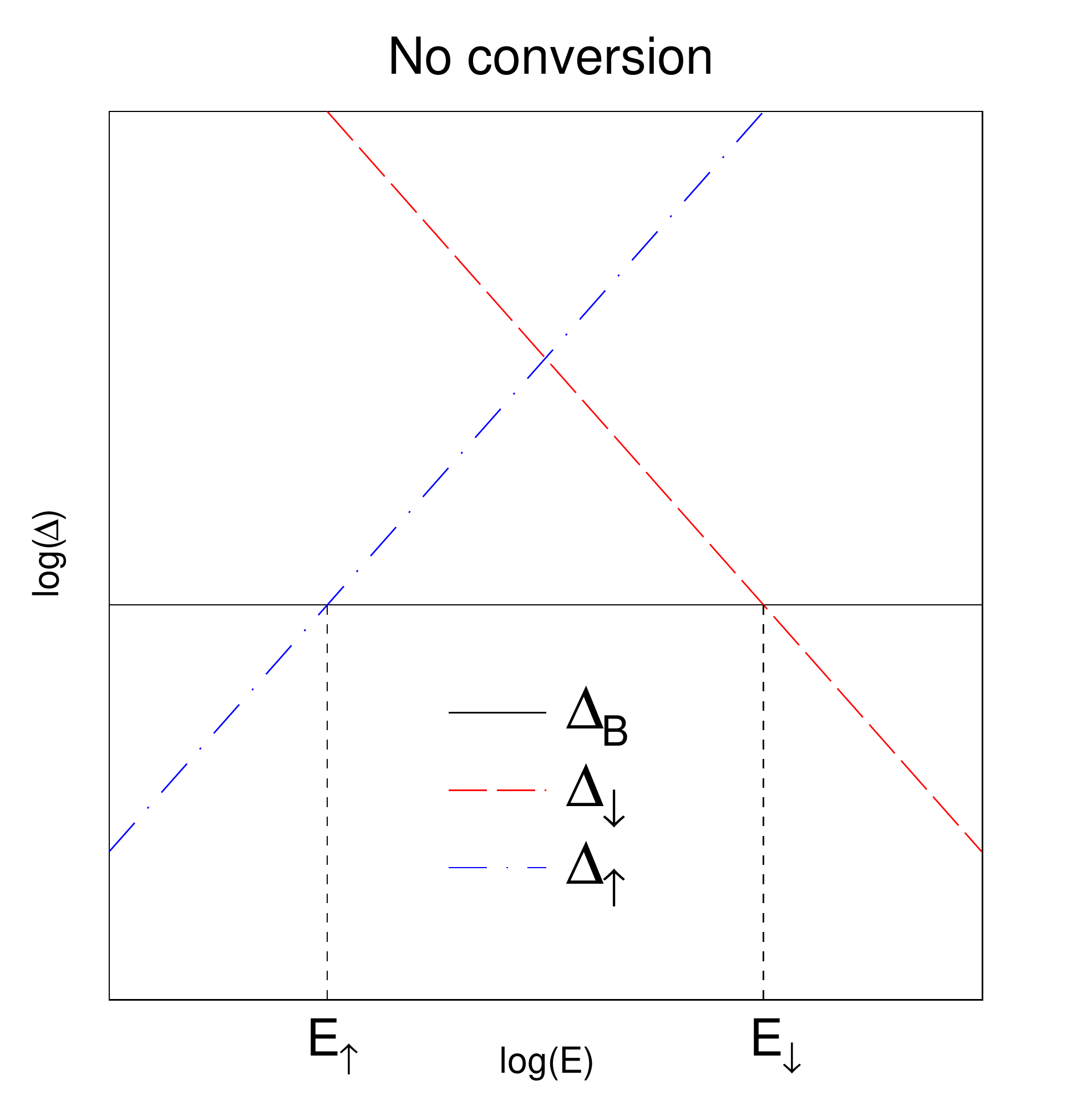}
   \caption{\label{fig:regimes}
   Illustration of the different conversion regimes depending on the critical energies:
   $E_\downarrow < E_\uparrow$: efficient conversion between $E_\downarrow$ and $E_\uparrow$ (left panel),
   $E_\downarrow \simeq E_\uparrow$: resonant conversion (middle panel),
   $E_\downarrow > E_\uparrow$: no efficient conversion, the interval of energy where conversion  would be possible does not exist anymore (right panel).}
\end{figure}

We are interested in cases where the transition from photons to scalars is not suppressed. This happens when the prefactor $\frac{1}{1+\frac{(\Delta_{\rm pl}-\Delta_{\rm a})^2}{4\Delta_{\rm B}^2}}$ is close to unity.
The conversion is therefore efficient when we have that  $\Delta_{\rm B} >  \vert \Delta_{\rm pl}-\Delta_{\rm a}\vert /2 $. As $\Delta_{\rm pl}$ and $\Delta_{\rm a}$ depend on the energy $E$ of the photon, only a band of energies  satisfies this condition. More precisely, there is one term proportional to the energy,
\be \Delta_\uparrow = \frac{b^2E}{1+b^2}, \ee
and one inversely proportional
\be
 \Delta_\downarrow = \frac{\omega_{\rm pl}^2}{2E} - \frac{m^2}{2E(1+b^2)}.
\ee
We have neglected the $B_z$ component and put $a\ll 1$ to comply with the bounds on the variation of $\alpha$ in the recent past of the Universe.
These terms are represented on Fig.~\ref{fig:regimes} for three different configurations. We have denoted by $E_{\uparrow/\downarrow}$ the energy at which $\Delta_{\uparrow/\downarrow}$ crosses $\Delta_{\rm B}$. When $E_\downarrow < E_\uparrow$, an efficient conversion takes place in this energy range. If not, no efficient conversion occurs. In the middle panel of Fig.~\ref{fig:regimes} the case when $E_{\downarrow}\simeq E_{\uparrow}$ (resonant conversion) is also illustrated.

We have  that
\begin{eqnarray}
E_{\downarrow} & = & \frac{|m^2-\omega_{\rm pl}^2(1+b^2)|\Lambda}{8B\sqrt{1+b^2}}\label{eq:edown} \\
E_{\uparrow} & = & \frac{4B}{\Lambda}\frac{\sqrt{1+b^2}}{b^2}.
\end{eqnarray}
 For $b = 0$, $E_\downarrow$ is equal to the critical energy in the ALP case above which the mixing between photons and ALP's is efficient.
The $\Delta_{\uparrow}$ and $\Delta_{\downarrow}$ terms cross at $E_\times = \sqrt{E_{\uparrow}E_{\downarrow}}$ .
We can now  find the range of parameters $m$ and $b$ for which an efficient conversion occurs, that is to say $E_{\downarrow} \ll E_{\uparrow}$.
For this, we solve $E_{\downarrow} = E_{\uparrow}$ for $b$ and get
\be
b_\times^2= \frac{m^2-\omega_{pl}^2 + \frac{32B^2}{\Lambda^2} \pm \sqrt{D}}{2\omega^2_{pl}}
\ee
where we have defined
\be
D= \left  (m^2-\omega_{pl}^2 + \frac{32B^2}{\Lambda^2} \right )^2 + 128\frac{\omega_{pl}^2B^2}{\Lambda^2}
\ee
Two cases need to be distinguished. When $ m^2<\omega_{pl}^2 (1+b^2)$, we have that $b_\times$ is unique and equal to
\be
b_0^2 \sim \frac{32B^2}{\Lambda^2 \omega_{pl}^2}
\ee
and an efficient conversion is only possible when $b<b_0$. Here the conversion is only efficient in an interval in energy between $E_\downarrow$ and $E_\uparrow$ and this is only possible for small values of the ratio $b=B/M^2\le b_0$. This is a new case compared to the usual ALP situation as the conversion there only involves one threshold energy and it becomes efficient
 above this threshold.
When $ m^2>\omega_{pl}^2 (1+b^2)$, we have two solutions for $b_\times$ given by
\be
b_{1}^2 \sim \frac{32B^2}{\Lambda^2 m^2},\ \ b_2^2= \left  (\frac{m^2}{\omega_{pl}^2}-1 \right ) \left (1-\frac{32B^2}{\Lambda m^2}\right )
\ee
In this case, when $b<b_1$ the conversion is efficient while for larger values of $b$ it is only efficient in the narrow band $b_2^2 <b^2<\frac{m^2}{\omega_{pl}^2}-1$.

\begin{figure}[h]
\includegraphics[width = 0.45\columnwidth]{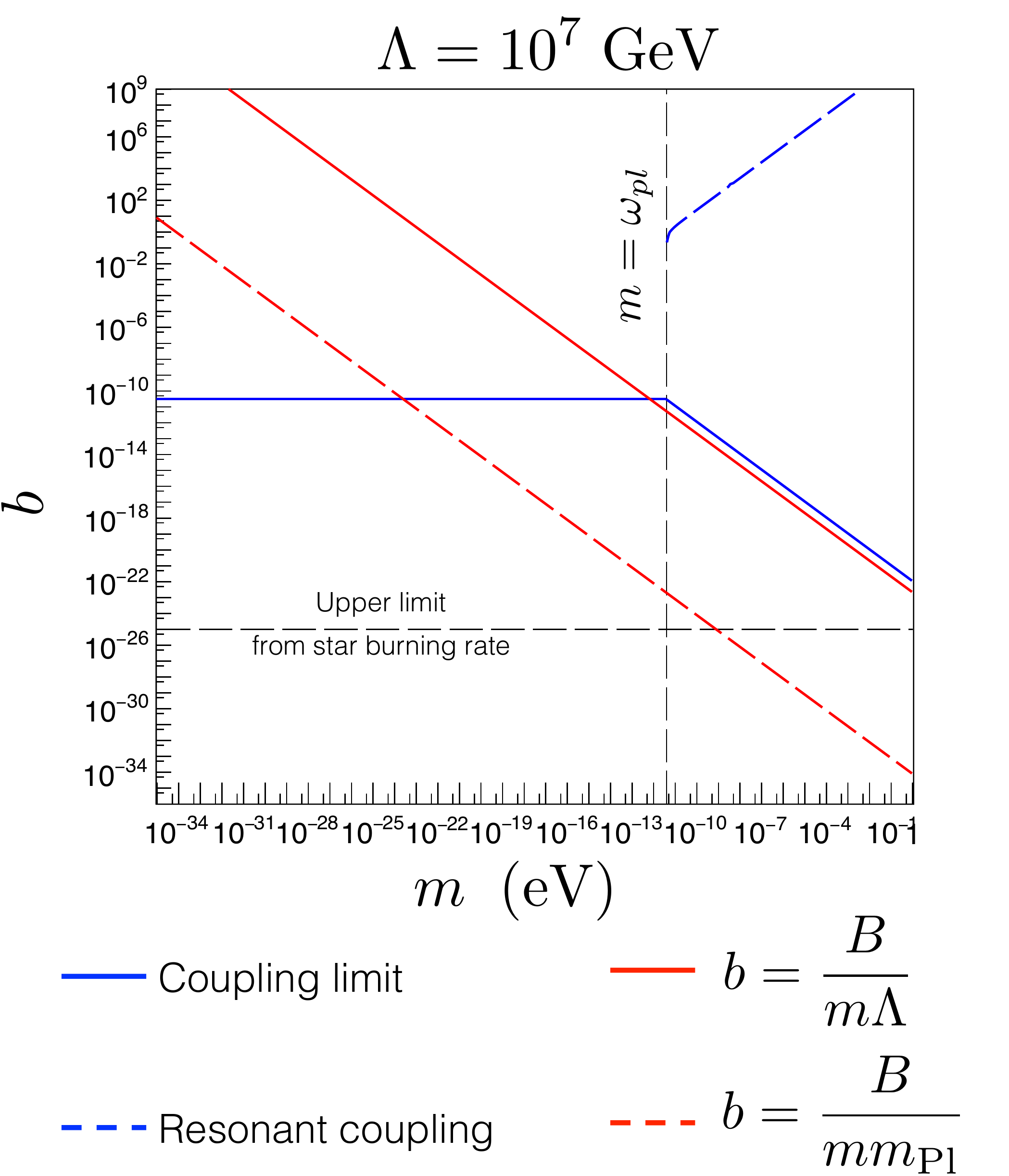}
\includegraphics[width = 0.45\columnwidth]{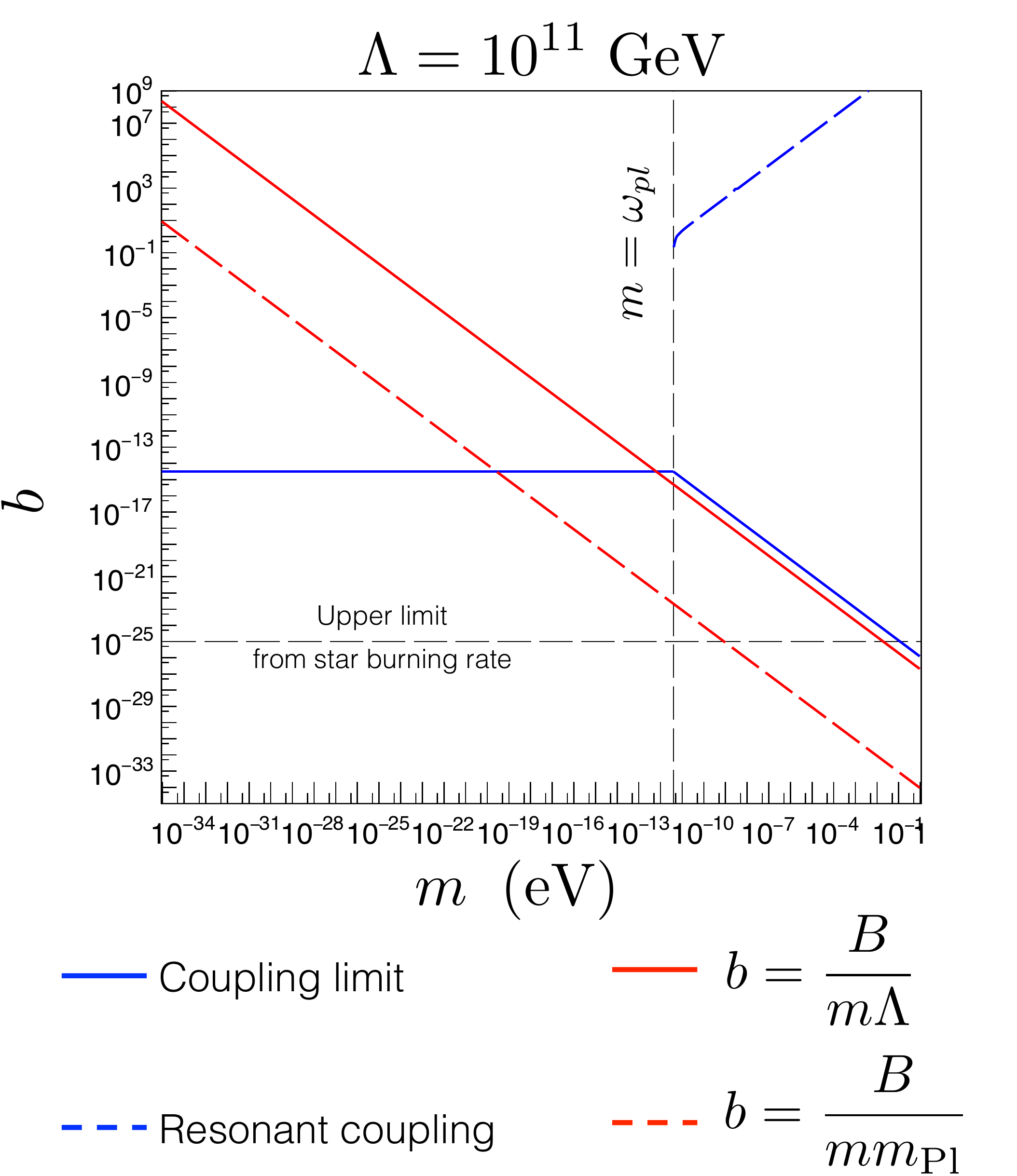}
\caption{\label{fig:b_vs_m}
Different regimes in the parameter space $b$ versus $m$ for values of $\Lambda=10^7 \rm \; GeV$ and  $\Lambda=10^{11}\;\rm GeV$ (right), see text for details.
}
\end{figure}

In some model realisations, it is possible to relate $b$ and $m$ using either the ansatz $b=B/m\Lambda $ or for massive gravity models $b=B/m m_{\rm Pl}$. We have summarised the results using these relations in Fig.~\ref{fig:b_vs_m} where we have assumed values for $B$ and $n_e$ of $10\;\rm \mu G$ and $0.01\;\rm cm^{-3}$ respectively that are typical of galaxy clusters, and a conformal coupling scales $\Lambda=10^7$ GeV (left) and $\Lambda = 10^{11}$ GeV (right). In Fig.~\ref{fig:b_vs_m}, the blue plain curves separate the two domains, where above this limit, no efficient conversion occurs. The blue dashed line is the region where resonant coupling happen, when $b^2 \sim \frac{m^2}{\omega_{pl}^2}-1$, in which case $E_{\downarrow} \rightarrow 0$. This requires a big fine-tuning of $b$ since the width is very narrow.  The red lines are just two examples of specific cases for $b$. For both panels of Fig.~\ref{fig:b_vs_m}, the red plain curve relates $b$ and $m$ using the scale $\Lambda$ and the red dashed curve using $m_{\rm Pl}$. Finally we have represented the horizontal like $b_{\rm {lim}}=B/M_{\rm lim} \sim 10^{-25}$ where $M_{\rm lim}= 346$ MeV is the limiting value allowed by the burning rate of stars. Only values of $b\le b_{\rm lim}$ are allowed when the constraint on $M$ is density independent. When $M$ and $\Lambda$ are related using $M^2=m\Lambda$, it appears that only a small mass window is accessible, just below the $m=\omega_{pl}$ line. If that constraint is relaxed, the mass range where signal is present lies somewhere between $\sim 10^{-25}$ eV and $10^{-13}$ eV for $\Lambda=10^7$ GeV, and between $\sim 10^{-20}$ eV and $10^{-13}$ eV for $\Lambda=10^{11}$ GeV. So considering larger conformal scales should narrow down the accessible mass window.

Note that when the relation $b=B/m m_{\rm Pl}$ is considered, the comparison between our potential limits and the ones from the star burning rate should be taken with caution, the scale $M$ should depend on the density because of screening effects. When the suppression scale $M$ depends on the density, such as for K-mouflage models, the limiting behaviour provided by the stellar burning constraint can be relaxed in a model dependent way. Indeed the density inside clusters is typically $\rho\sim 10^{-24} \ {\rm g\;cm^{-3}}$ whereas it is typically $\rho \sim 100 \ {\rm g\;cm^{-3}}$ in stars and even a mild dependence on the density may render the star constraint weaker than the bound obtained from galaxy clusters. The analysis of this dependence is left for future work.

\begin{figure}[h]
\includegraphics[width = 0.45\columnwidth]{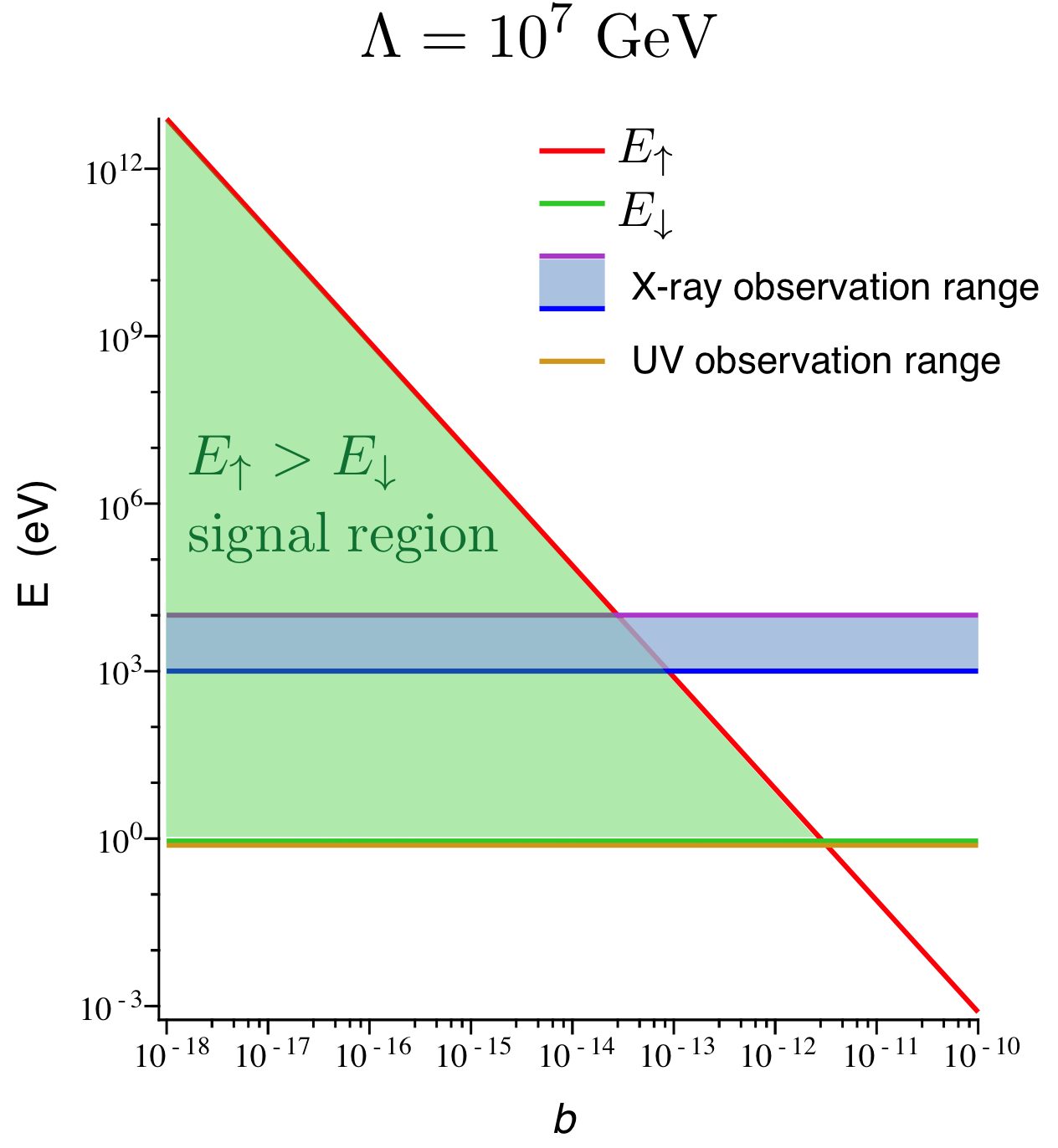}
\includegraphics[width = 0.45\columnwidth]{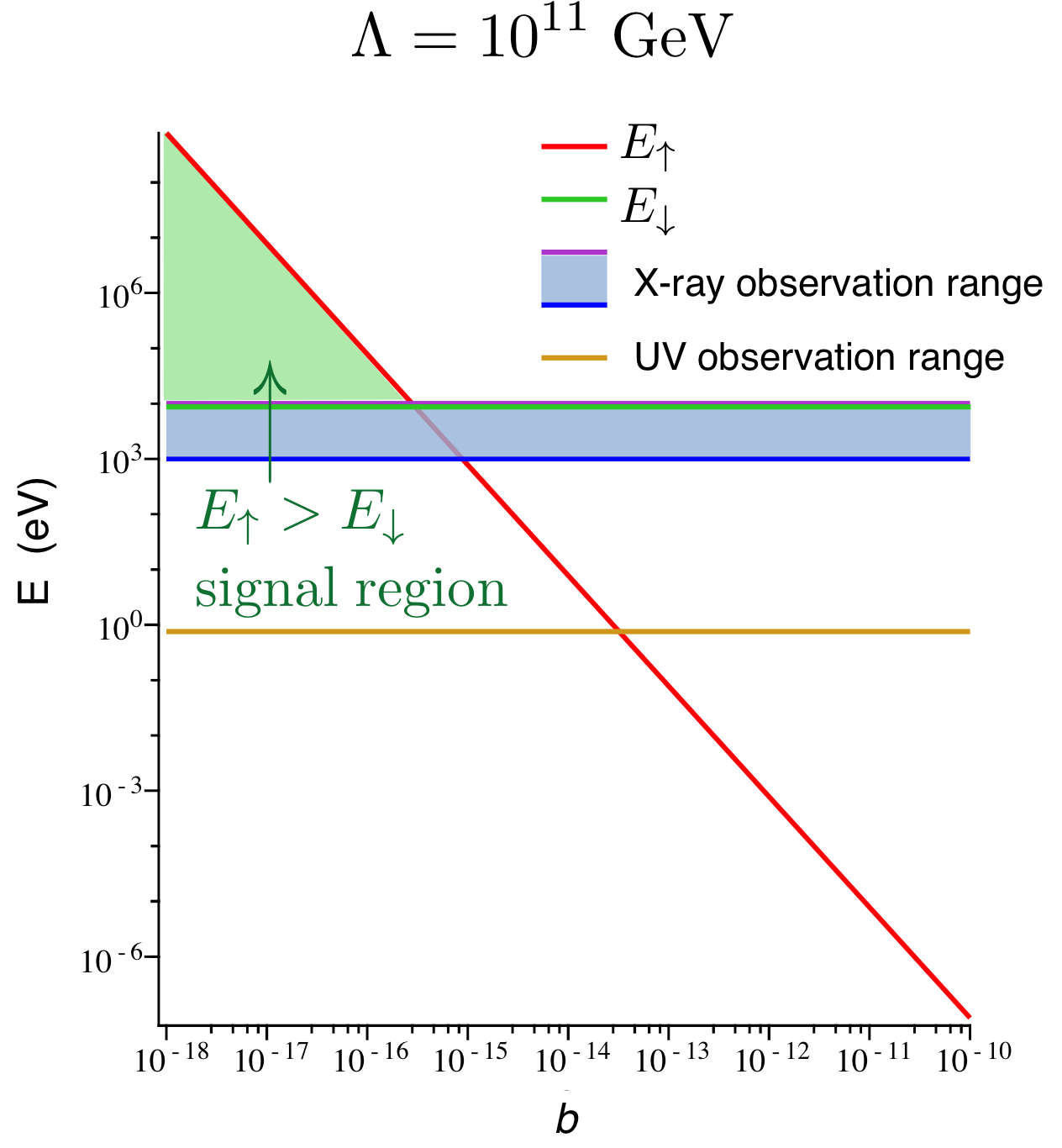}
\caption{\label{fig:3} Illustration of the accessible values for $b$, for $\Lambda=10^7$ GeV (left) and  $\Lambda=10^{11}$ GeV (right).}
\end{figure}

In the following, two energy bands will be considered for the photons: X rays between 1 keV and 7 keV, and UV around 1 eV. From the above analysis of the energy dependence of the signal, it is already possible to estimate the ranges in $\Lambda$ and $b$ that will be accessible to this study. This is illustrated in Fig.~\ref{fig:3}, where $\Lambda$ is fixed at $10^7$ GeV (left panel) and $10^{11}$ GeV (right panel). The region where an irregularity signal is potentially observable ($E_\downarrow < E_\uparrow$) is highlighted in green. The two energy ranges for photon observation are displayed as horizontal zones. From the left panel of Fig.~\ref{fig:3} we infer that (if high enough) the signal will be accessible for $b$ lower than $\sim 10^{-11}$ in UV and $\sim 10^{-13}$ in X rays. Increasing $\Lambda$ up to $10^{11}$ GeV narrows the signal region and make it inaccessible to both energy ranges. It is therefore expected that it will not be possible to reach higher $\Lambda$ values.

\begin{figure}[h]
\includegraphics[width = 0.45\columnwidth]{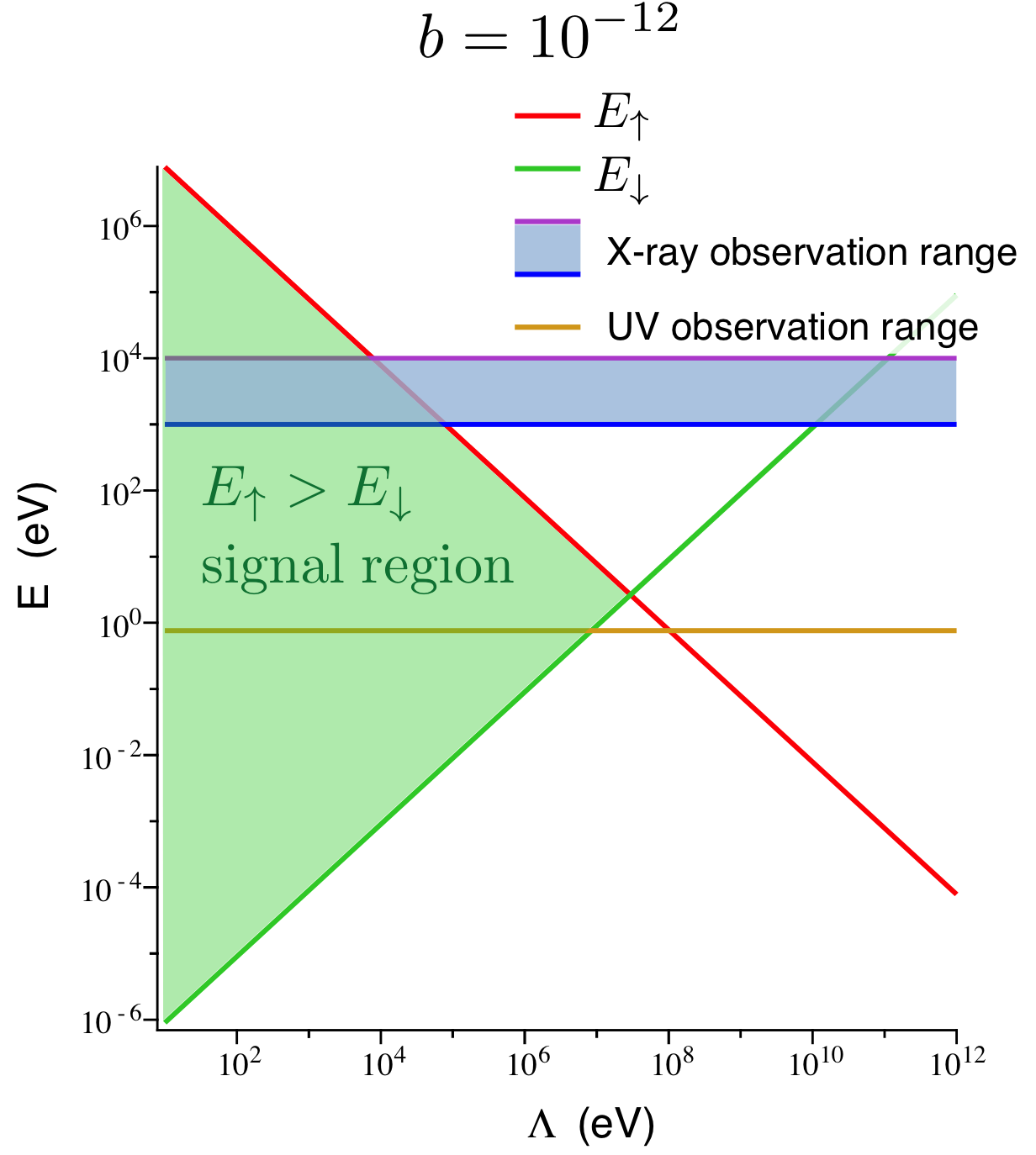}
\includegraphics[width = 0.45\columnwidth]{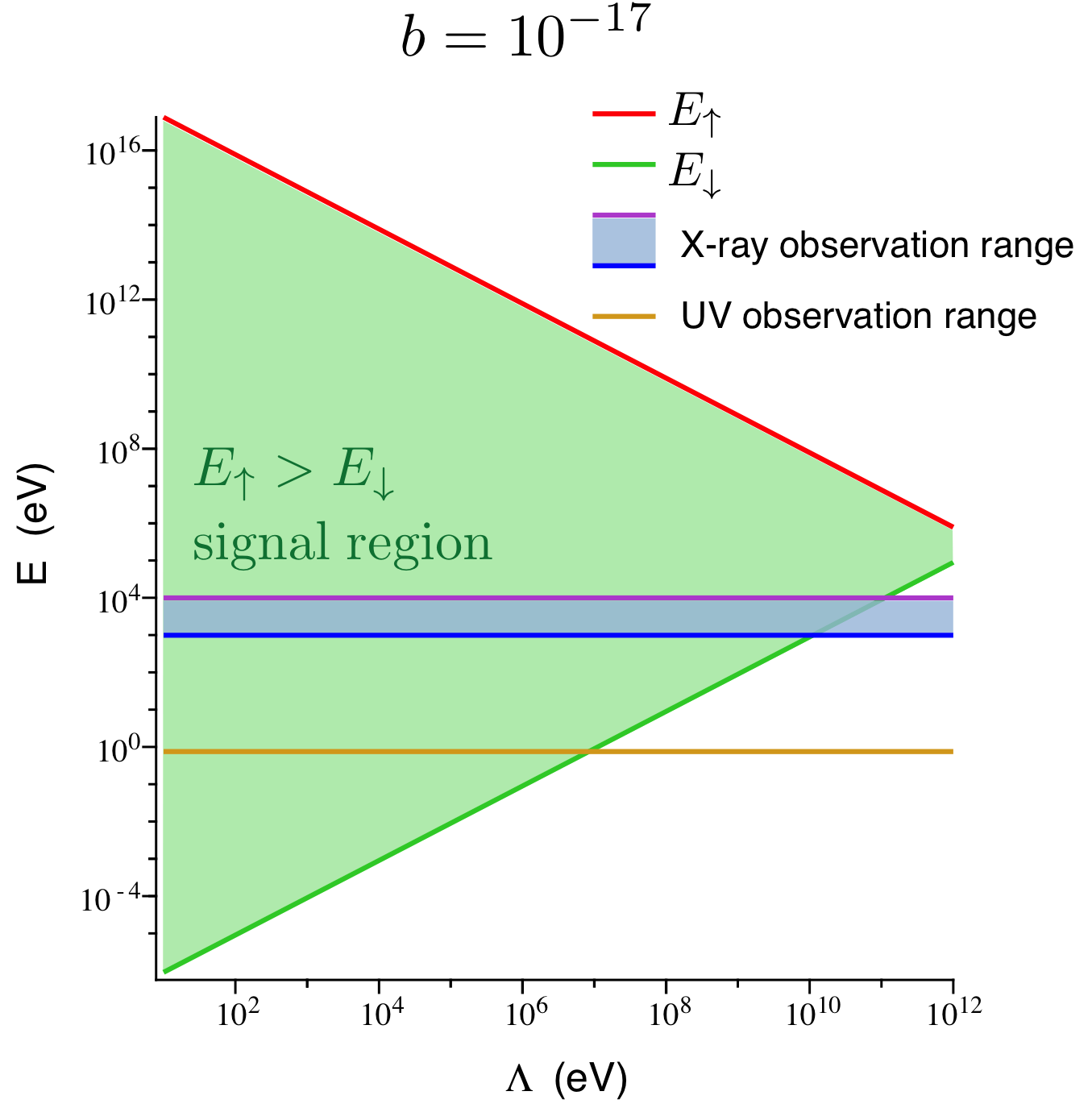}
\caption{\label{fig:4} Illustration of the accessible values for $\Lambda$, for $b=10^{-12}$ (left) and  $b=10^{-17}$  (right).}
\end{figure}

In a similar fashion, one can fix values of $b$ to draw accessible regions in the  $(E,\Lambda)$ plane. This is illustrated in Fig.~\ref{fig:4}, where the signal regions are drawn for two values of $b$, $10^{-12}$ (left) and $10^{-17}$ (right). From these figures, it appears that in the case when $b=10^{-12}$, accessible values for $\Lambda$ are below $10^5$ GeV and $10^{7}$ GeV for X rays and UV respectively. When $b$ is reduced to $10^{-17}$, these values shift to $10^{11}$ and $10^7$ respectively.

\subsection{Laboratory experiments}

The disformal coupling to radiation has effects on the the propagation of laser beams in optical cavities in the presence of a transverse magnetic field.
Indeed, the mixing between the photons and the scalar field implies that the polarisation of an initially linearly polarised beam rotates and becomes slightly elliptical.
The strongest constraints are  given by the PVLAS experiment \cite{Zavattini:2007ee} which cannot test models when $\Lambda \gtrsim 10^5$ GeV. Better constraints are obtained from the light shining through a wall experiment at DESY \cite{Ehret:2010mh}
where the probability than an initial laser beam is regenerated on the other side of a wall due to the crossing of the wall by the scalar field. In this case, the experiment cannot constrain models with $\Lambda\gtrsim 10^7$ GeV. For lower values of $\Lambda$, the masses of the scalar field are extremely constrained when $M^2= m_{\rm Pl} m$, with $m\lesssim H_0 \sim 10^{-33}$ eV  \cite{Brax:2012ie}. All these laboratory tests have been performed in densities $\rho\sim 10^{-14} \ {\rm g.cm^{-3}}$.
This should be compared with  the available mass range  obtained using $b<b_0\sim \frac{4\sqrt 2 B}{\Lambda m_{\rm Pl}}$ and $M^2=m\,m_{\rm Pl}$ leading to
\be
m \gtrsim m_\star=\frac{\Lambda \omega_{pl}}{m_{\rm Pl}}
\ee
which can be as low as $10^{-31}$ GeV.
There is also the astrophysical limit $b<b_{\rm lim}=\frac{B}{M^2_{\rm lim}}$
\be
m\gtrsim \frac{M^2_{\rm lim}}{m_{\rm Pl}}
\ee
The latter prevents to probe masses of
order $H_0$ and in fact $m\gtrsim 10^{-19}$ GeV.  Hence, laboratory constraints imply that astrophysical tests of the disformal coupling can yield new result only for  large values of $\Lambda\gtrsim 10^7$ GeV and masses $m\gtrsim \sup({m_\star,m_{\rm lim}})$. This can be seen in Fig~\ref{fig:b_vs_m} where $m_{\rm lim}> m_\star$ for $\Lambda=10^{11}$ GeV. In summary, the limits that we will derive from galaxy clusters cannot probe the very small mass range where $m\sim H_0$ but larger masses which are still very small compared to the masses of known particles.

\section{Propagation in turbulent magnetic fields}
\label{sec:turbulent}

The case of a homogeneous magnetic field that was considered in the previous section is a simplified picture that is not representative of astrophysical magnetic fields. For instance, measurements of the Faraday rotation of the polarization of synchrotron emission from radiogalaxies embedded in galaxy clusters showed that magnetic fields in clusters are usually turbulent on various spatial scales~\cite{2003A&A...412..373V}, following a Kolmogorov power spectrum ($P(k) \propto k^{-5/3}$) up to scales of tens of kpc~\cite{2001A&A...379..807G,2010A&A...513A..30B,2011A&A...529A..13K}. The Milky Way also hosts a turbulent magnetic field, in addition to the large scale ordered component (see~\cite{Jansson:2012pc} for a recent model), that is found to a have a Kolmogorov power spectrum on scales from 0.3 pc to 100 pc~\cite{1996ApJ...458..194M}. Notice that the interaction between the scalar field and photons do not affect these measurements, which are in the radio band.

In the following, the turbulence of the magnetic field is described as in~\cite{1999ApJ...520..204G} by summing over a large number of plane waves having randomly chosen propagation directions. The plane waves have different wavenumbers and corresponding amplitudes so as to describe the power spectrum assumed for the turbulence. This method has the advantage of generating a three-dimensional turbulent magnetic field that can be computed at any point in space coordinates, based only on a random set of angles for the propagation direction and phase of each plane wave. While one realization of the turbulent magnetic field corresponds to a random choice of every angles, it is obvious that the exact 3D realization of the turbulent magnetic field cannot be measured in nature and thus remains unknown. In the description of the turbulence with a Kolmogorov power spectrum, the fluctuations on the smallest spatial scales are less important than the fluctuations at the highest scale and therefore become irrelevant as the scales decrease. The line of sight for the propagation in the turbulent magnetic field of the system, initially consisting of photons, is then divided in small domains such that in each domain the magnetic field is assumed to be homogeneous, such that equation~\ref{eq:equa_motion} is easily integrated.

The initial beam is conservatively assumed to be unpolarized. This is accounted for in the density matrix formalism:
\be
\rho\;\;=\;\;\left ( \begin{array}{c} A_x \\ A_y \\ i\phi \end{array} \right ) \otimes \left ( \begin{array}{ccc} A_x & A_y & i\phi \end{array} \right )^\star\;\;,
\ee
and an initial density matrix $\rho_0 = \text{diag}(\nicefrac{1}{2},\nicefrac{1}{2},0)$. After propagation in every magnetic domains, the density matrix is computed recursively as
\be
\rho_{k+1} = e^{-i\mathcal{M}_k \,s}\cdot \rho_{k} \cdot e^{i\mathcal{M}^\dag_k \,s}\;\;,
\label{eq:recursive}
\ee
where $s$ is the size of the domain crossed by the beam and $\mathcal{M}_k$ is the mixing matrix given by Eq.~\ref{eq:mixing_matrix}. The 3-dimensional magnetic field used in the mixing matrix is computed from the turbulence model randomly generated and is assumed to be homogeneous in each domain. The value and direction of the magnetic field between the domains are related through the power spectrum assumed in the turbulence model.

It has been shown in~\cite{2012PhRvD..86d3005W} for the ALP case, that the photon survival probability after crossing various magnetic domains, that can be computed with Eq.~\ref{eq:recursive}, has a very complex energy dependence around the critical energy. This effect is due to the turbulent nature of the magnetic field. The irregular pattern in the photon survival probability around the critical energy depends on the exact realization of its turbulence. In the present case where the mixing only occurs for energies in the energy range between $E_\downarrow$ and $E_\uparrow$, spectral irregularities are present around both critical energies. This signature is similar to the ALP case with the addition of a new parameter $b$, that modifies the irregularities pattern when $b\sim b_\times$. For this reason, constraints that have already been obtained on ALPs with the spectral irregularities signature (see~\cite{2013ApJ...772...44W,2013PhRvD..88j2003A}) cannot be translated directly to constrain disformal couplings. As explained in Sec.~\ref{Introduction}, we are mainly interested in constraining the very low mass domain for the scalar field. This can be achieved with the addition of the plasma term in the mixing matrix, as it was considered in~\cite{2013ApJ...772...44W}, so that the constraints are valid for all masses $m \ll \omega_{pl} \sim 10^{-11} \rm\,eV$ (see Eq.~\ref{eq:edown}).

\begin{figure}[h]
\includegraphics[width = \columnwidth]{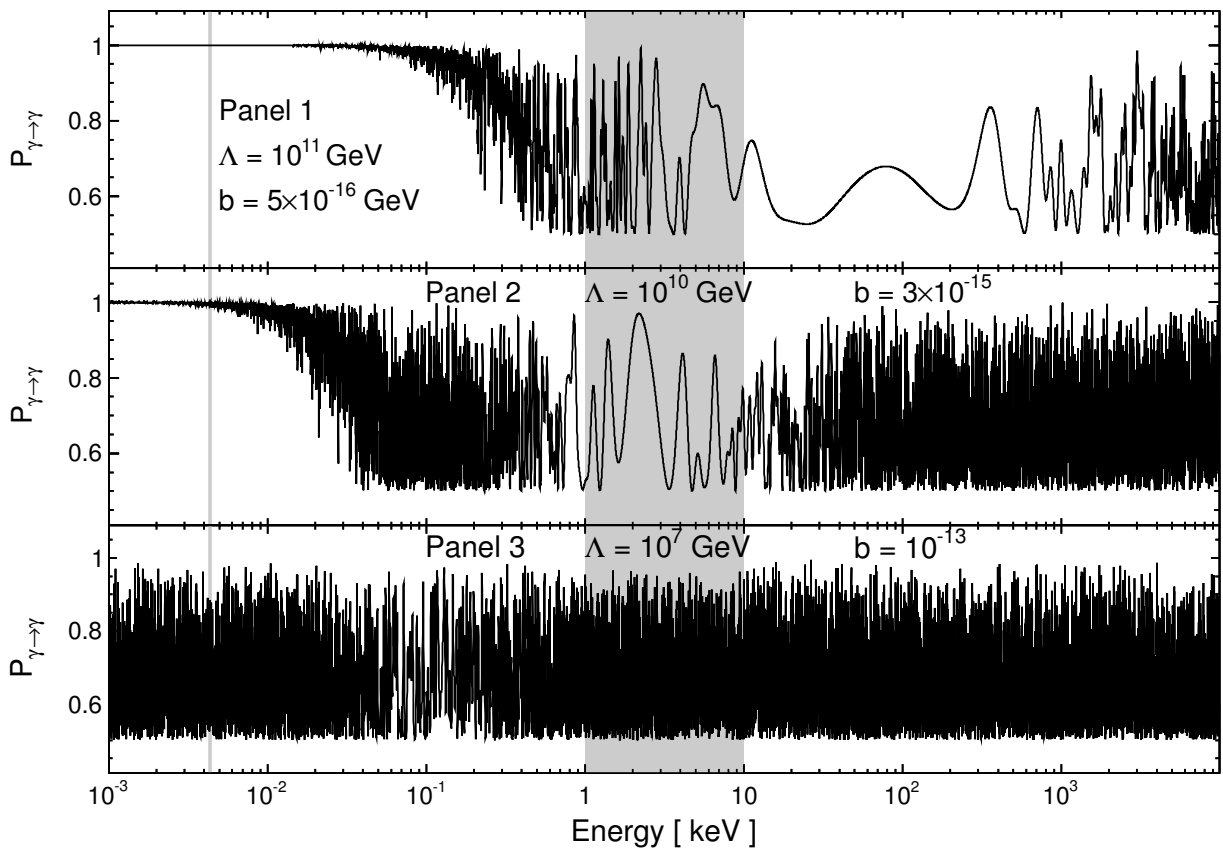}
\caption{Photon survival probability for three values of $b$, assuming propagation in one realization of the turbulent magnetic field model of Sec~\ref{sec:observations}. The grey area shows the energy range sensitive to X-ray observations and the grey line the energy domain probed by the UV spectroscopy of PKS 2155-304.}
\label{fig:turbulence}
\end{figure}

An example of the photon survival probability from an initial unpolarized beam of pure photons is shown on Fig~\ref{fig:turbulence} for three values of $b$ and $\Lambda$ assuming propagation in a random realization of a magnetic field typical of galaxy clusters with $B = 1\,\mu$G and $n_e = 0.01\rm\,cm^{-3}$. As shown in section~\ref{sec:critical}, when $b$ is increased, $E_\uparrow$ decreases and the energy range for the strong mixing is reduced. Nevertheless, the expression of $E_\uparrow$ shows that this could be in principal compensated by lower values of $\Lambda$. For lower values of $\Lambda$, the oscillation length of the mixing is decreased and the spectral irregularities are tighter. This is shown on the second panel of Fig.~\ref{fig:turbulence} where $\Lambda = 10^{10}\rm\,GeV$ as compared to the first panel where $\Lambda = 10^{11}\rm\,GeV$. For this reason, two methods are proposed in the following. The first method has been presented in~\cite{2013ApJ...772...44W} and is based on X-rays of the Hydra A galaxy cluster. Due to the limited energy resolution of the instrument, of the order of $5\%$, this method can probe spectral irregularities at high $\Lambda$, and corresponding low $b$. This case is outlined on panel 1 and 2 of Fig.~\ref{fig:turbulence} where the gray area shows the sensitive energy range. In order to identify spectral irregularities on small energy scales, a spectral resolution of the order of $10^{-4}$ is required. A second method based on high resolution spectroscopy of a BL Lac object, PKS 2155-304, is then used to obtain constraints on lower values of $\Lambda$ but higher values of $b$ (see panel 3 of Fig.~\ref{fig:turbulence}).

\section{Constraints from X-ray observations}
\label{sec:observations}

Because of typical values for the electron density and magnetic fields in galaxy clusters, the critical energy $E_\downarrow$ is expected to be of order of a keV. Consequently, spectral irregularities due to the conformal mixing are expected to occur in this energy band, as shown in Fig.~\ref{fig:turbulence}. Spectral irregularities should then be searched in the spectrum of a bright X-ray source embedded in a galaxy clusters with a strong magnetic field. As explained in~\cite{2013ApJ...772...44W}, the non-thermal emission from the central AGN of Hydra A is a good target for the present search because of the strong magnetic field in its cluster that has been extensively studied in the literature~\cite{1993ApJ...416..554T,2003A&A...412..373V,2005A&A...434...67V,2008MNRAS.391..521L,2011A&A...529A..13K}.

\subsection{Magnetic fields in the Hydra A galaxy cluster}

The magnetic field in the Hydra A galaxy cluster can be probed with Faraday rotation maps from the radio emission of the lobes of the central AGN. To obtain a measurement of the magnetic field, a profile of the electron density is assumed. It has been estimated in~\cite{2005A&A...434...67V} by using the X-ray surface brightness measured by ROSAT in~\cite{1999ApJ...517..627M} with the method of~\cite{2004A&A...413...17P}. The electron density profile is parameterized as a function of the radius $r$ from the central source following~\cite{2011A&A...529A..13K}:
\be
n_e(r) = \left[n_{e1}^2\left(1+\left(\frac{r}{r_{c1}}\right)^2\right)^{-3\beta}+n_{e2}^2\left(1+\left(\frac{r}{r_{c2}}\right)^2\right)^{-3\beta}\right]^{1/2} \;\;,
\label{eq:profile}
\ee
with $n_{e1} = 0.056\rm\,cm^{-3}$, $n_{e2} = 0.0063\rm\,cm^{-3}$, $r_{c1} = 33.3\rm\,kpc$, $r_{c2}= 169\rm\,kpc$ and $\beta = 0.766$. The same electron density profile is used later in the simulations of the propagation of the photon/scalar system to compute the plasma term along the line of sight. Moreover, the r.m.s. of the turbulent magnetic field at a radius $r$ in the cluster is believed to scale with the electron density as $B(r)\propto n_e(r)^{\alpha_{B}}$~\cite{2001A&A...378..777D}.

The measurement of the magnetic field also depends on the geometry that is assumed for the orientation of the jets and lobes from which the Faraday rotation maps are produced. The angle of projection of the northern jet on the line of sight $\theta$ explains the depolarization asymmetry that is observed between the two lobes~\cite{1993ApJ...416..554T}. The uncertainty on this parameter still remains large, as $\theta$ is constrained in the range between $30^\circ$ and $60^\circ$. In the latest analysis~\cite{2011A&A...529A..13K}, a magnetic field strength $B_0$ at $r=0$ of $21\,\mu\rm G$ is found if $\theta = 30^{\circ}$, compared to $B_0 = 36\rm\,\mu G$ if $\theta = 45^\circ$ and $B_0 = 85\,\mu\rm G$ if $\theta = 60^\circ$. To be conservative, $B_0 = 21\,\mu\rm G$ is assumed in the following. In the same analysis, a most likely value of 1 for the scale parameter $\alpha_B$ is found. This value is also used in the present work. For the turbulence of the magnetic field, a Kolmogorov power spectrum is measured on spatial scales up to 10 kpc. For the simulations of the turbulent magnetic field performed in this study, a minimal scale of 1 kpc is assumed. As explained in~\cite{2012PhRvD..86d3005W}, spectral irregularities rapidly become irrelevant for lower scales because of the joint effect of the Kolmogorov power spectrum that suppress contributions on scales $s$ as $s^{-2/3}$~\cite{2007PhRvD..76b3001M} and the mixing efficiency that is suppressed as $s^{-1/2}$. A Kolmogorov power spectrum on scales between 1 kpc and 10 kpc is then used in the following for the  simulations of the turbulent magnetic field with the method described in Sec~\ref{sec:turbulent}.

\subsection{\textit{Chandra} observations of Hydra A}

Observations of the Hydra A galaxy cluster with \textit{Chandra}~\cite{2000SPIE.4012....2W} are analyzed following the method outlined in~\cite{2013ApJ...772...44W}. The source has been observed in 1999 and 2003~\cite{McNamara:2000kj,2009ApJ...707L..69K} with the ACIS instrument composed of two arrays of imaging CCDs that are sensitive to X-rays between 0.2 and 10 keV~\cite{1999NIMPA.436...40B}. The energy resolution of ACIS of 0.1 keV at 1.5 keV~\cite{2003SPIE.4851...28G} enables the search for spectral irregularities due to photon/scalar mixing. A total live-time of 238 ks is used in the analysis.

\begin{figure}[h]
\includegraphics[width = 0.5\columnwidth]{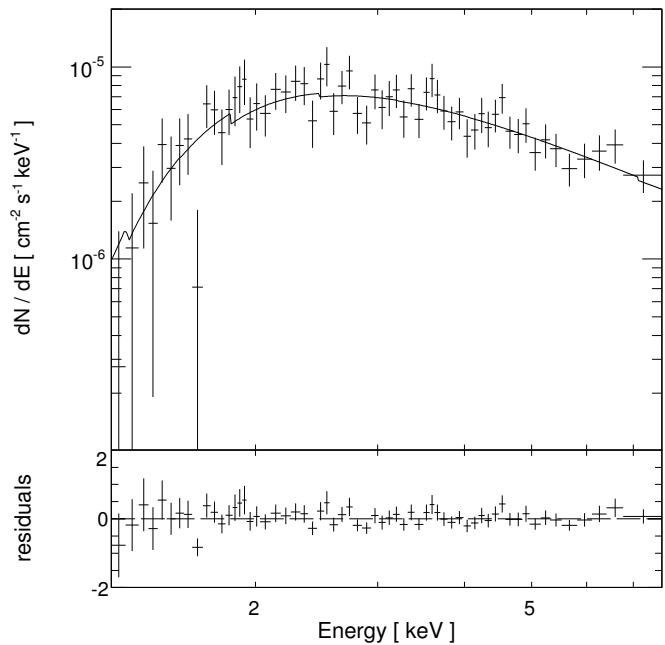}
\caption{X-ray spectrum of the central source in Hydra A galaxy cluster observed by \textit{Chandra}. Top panel: reconstructed spectrum with model (see text for details). Bottom panel: residuals to the model.}
\label{fig:spectrum}
\end{figure}

The spectral analysis, performed with the \texttt{XSPEC} package version 12.7.1, shows that the spectrum of the central source is well-modeled ($\chi^2/n_{\rm dof} = 48.59/58$) by a power-law with a photoelectric absorption term ($dN/dE=\phi_0 E^{-\Gamma}e^{-\tau}$) avec $\Gamma = 1.52\pm0.17$. The optical depth $\tau$ is due to photoelectric absorption and is described by the hydrogen column $N_{\rm H} = 2.54\pm0.33\times10^{22}\rm\,cm^{-2}$ as $\tau = N_{\rm H}\sigma(E(1+z))$ where $\sigma(E)$ is the photoelectric cross-section as a function of the energy. As the absorption is mostly due to the intracluster medium of Hydra, the redshift of the source $z = 0.0538$ is taken into account. The spectrum is shown on Fig.~\ref{fig:spectrum} together with the model and the residuals of the fit. Because of the strong photoelectric absorption, the central source is not visible above 1 keV and the search for spectral irregularities is therefore restricted to energies higher than this threshold.

\subsection{Statistical analysis of the irregularity level in the X-ray spectrum}
\label{sec:statX}

In order to estimate if some spectral irregularities induced by the mixing between photons and the scalar field can be accommodated by the data on Hydra A, they are fitted to the spectrum obtained in the previous section. Assuming that the intrinsic spectrum of the central source is well modeled by a power law with an absorption term (see $dN/dE$ above), the data are fitted with this spectral shape modulated by the irregularities due to the scalar-photon coupling. The spectral shape fitted is then $dN/dE\times P_{\gamma\rightarrow\gamma}$ where $P_{\gamma\rightarrow\gamma}$ is obtained by the propagation of an unpolarized photon beam in a random realization of the magnetic field. When $\Lambda\to\infty$, $P_{\gamma\rightarrow\gamma} = 1$ and the initial spectral shape $dN/dE$ is retrieved. The assumption of this particular spectral shape does not bias the result because a power law is the simplest spectral shape (in this context, the most regular) that could be used to describe the data. Assuming a more complicated spectral shape could yield a better constraint as the model would provide additional degrees of freedom and potentially accommodate artificially statistical fluctuations, thus excluding spectral irregularities more efficiently.
As in~\cite{2013ApJ...772...44W}, the goodness of fit is computed with a likelihood estimator $\mathcal{L}$ that assumes that the number of events in each reconstructed energy bin $i$ follows a Poisson distribution:
\be
\mathcal{L} = \prod_{i=1}^{N}\frac{[t_s(m_i+b_i)]^{S_i}e^{-t_s(m_i+b_i)}}{S_i!}\frac{(t_bb_i)^{B_i}e^{-t_bb_i}}{B_i!}\;\;,
\ee
 where in each reconstructed energy bin $i$,  $S_i$ is the number of measured events from the signal region defined by a circle of $1^{\prime\prime}$ around the central source and $B_i$ is the number of measured events from the background region defined as an annulus between $1^{\prime\prime}$ and $2.5^{\prime\prime}$ around the same position. The exposure rates of the two regions, corrected from their surface ratio, is written $t_s$ and $t_b$ for respectively the signal and the background region. The expected signal photon rate $m_i$ is predicted from the assumed spectral shape, convolved by the instrumental response functions:
\be
m_i = \int_{\mathrm{bin}~i}\int_0^\infty\frac{\mathrm{d}N}{\mathrm{d}E}\times P_{\gamma\rightarrow\gamma}(E) A(E)P(\tilde{E}|E)\mathrm{d}E\mathrm{d}\tilde{E} \;\;,
\ee
where $A(E)$ is the true energy effective area and $P(\tilde{E}|E)$ the migration matrix from true energies to reconstructed energies. The expected background photon rate is estimated from the measurement in the background region by requiring
\be
\frac{\partial\ln\mathcal{L}}{\partial b_i} = 0 \;\;,
\ee
giving the most likely value for $b_i$.

For each set of parameters $(b,\Lambda)$, many simulations (1000) of the propagation of the system in the Hydra A magnetic field are performed. Each simulation corresponds to different realizations of the turbulent magnetic field and therefore gives different spectral irregularities patterns. To account for the lack of knowledge on the exact realization of the turbulent magnetic field, a likelihood ratio test is computed where the configuration of the magnetic field is considered as a nuisance parameter~\cite{2005NIMPA.551..493R}:
\be
\lambda(b,\Lambda) = \frac{\sup\limits_{\theta} \mathcal{L}(b,\Lambda, \theta)}{\sup\limits_{b,\Lambda, \theta}\mathcal{L}(b,\Lambda, \theta)} \;\;.\label{eq:maxL}
\ee
The log-likelihood ratio test  $-2\ln\lambda$ follows a $\chi^2$ distribution with two degrees of freedom~\cite{2005NIMPA.551..493R} and the exclusion obtained on $(b,\Lambda)$ at 95\% C.L. corresponds to the set of parameters $(b,\Lambda)$ for which $-2\ln\lambda(b,\Lambda) > 6.17$.  The obtained constraints on the model parameters are presented in Sec.~\ref{sec:constraints}.

\section{UV spectroscopy of PKS 2155-304}
\label{sec:statUV}

Because of the expression of $E_\uparrow$, it is possible to probe higher values of $b$ than what can be achieved with X-ray observations by searching for spectral irregularities  at lower energies. This is rendered possible  by looking at lower values of $\Lambda$, which corresponds to lower values of $E_\downarrow$. The spectroscopy of a bright optical-UV point-like souce embedded in a galaxy cluster is used to obtain the constraints. In this study, we use the spectroscopy of PKS 2155-304, which is a BL Lac object around which a galaxy cluster is observed. This object has already been used in~\cite{2013PhRvD..88j2003A} to search for ALPs via spectral irregularities at TeV energies. Here, we take over the description of the magnetic field in the galaxy cluster of PKS 2155-304 that was used in~\cite{2013PhRvD..88j2003A}, where the magnetic field was conservatively described with a strength of 1 $\mu$G and a Kolmogorov turbulence power spectrum on spatial scales between 1 kpc and 10 kpc. A typical electron density in galaxy clusters of 0.01 cm$^{-3}$ is assumed.

\begin{figure}[h]
\includegraphics[width = 0.8\columnwidth]{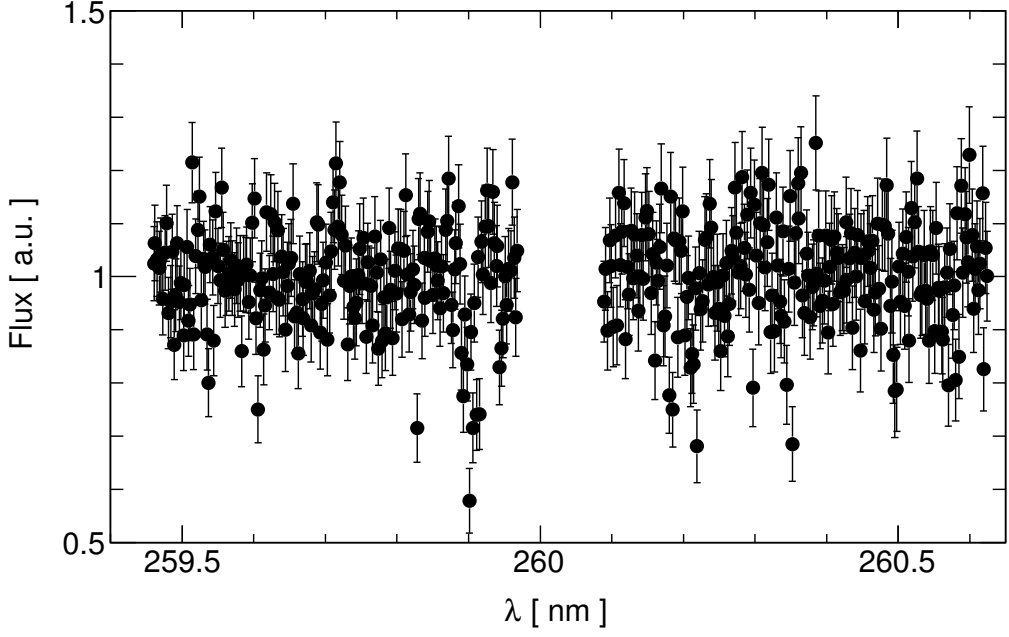}
\caption{Spectrum of PKS 2155-304 measured by HRS with the ECH-B echelon grating.}
\label{fig:spectrum_hubble}
\end{figure}

The optical and UV spectra of PKS 2155-304 have been measured by the HRS instrument on-board  \textit{Hubble}. The ECH-B echelon grating was chosen for its high resolving power of 1:80000 that is required to identify spectral irregularities occurring on the smallest energy scales. Observations were carried out in May 1993, before the first servicing mission that installed the COSTAR optics. The spectrum measured  between 259.4 nm  and 260.6 nm with the exclusion of a feature between 259.97 and 260.07 is shown in figure~\ref{fig:spectrum_hubble}. The spectrum is shown normalized to its average value of $7.4\times10^{-14}\rm\,erg\,cm^{-2}s^{-1}\angstrom^{-1}$. The fit of a constant to the data gives a $\chi^2$ of 718.1 for 416 degrees of freedom, indicating that the spectrum is affected by some spectral features. The method used to search for spectral irregularities induced by the coupling between photons and the scalar field is similar to the one used in the \textit{Chandra} analysis. Here, an underlying constant spectral shape is assumed, and a $\chi^2$ estimator is used to quantify the goodness of fit. While the presence of some spectral features in the measured spectrum may decrease the sensitivity to spectral irregularities, this method is conservative when deriving constraints.

\section{Constraints on the model parameters}
\label{sec:constraints}

\begin{figure}[h]
\includegraphics[width = 0.49\columnwidth]{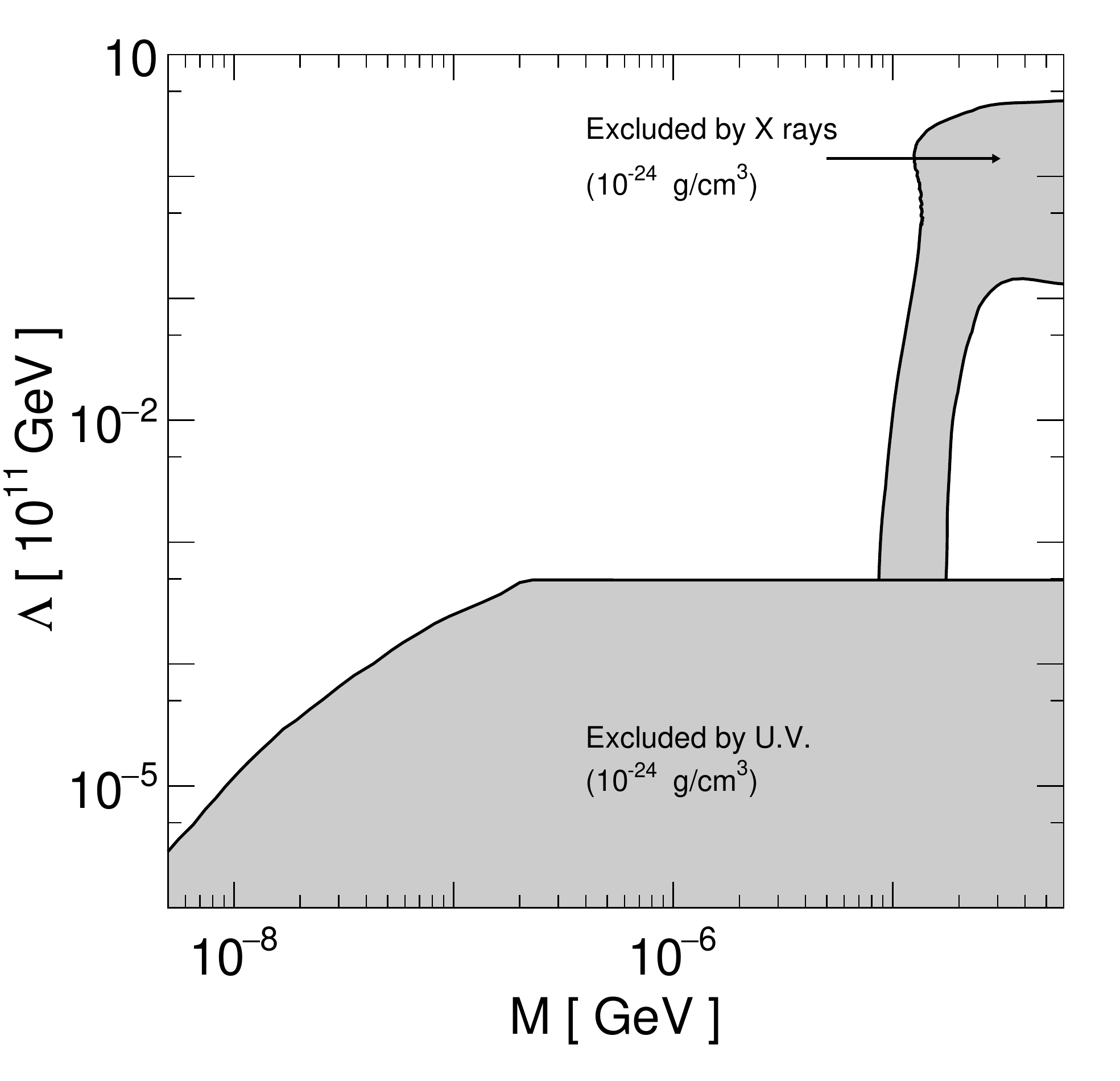}
\includegraphics[width = 0.49\columnwidth]{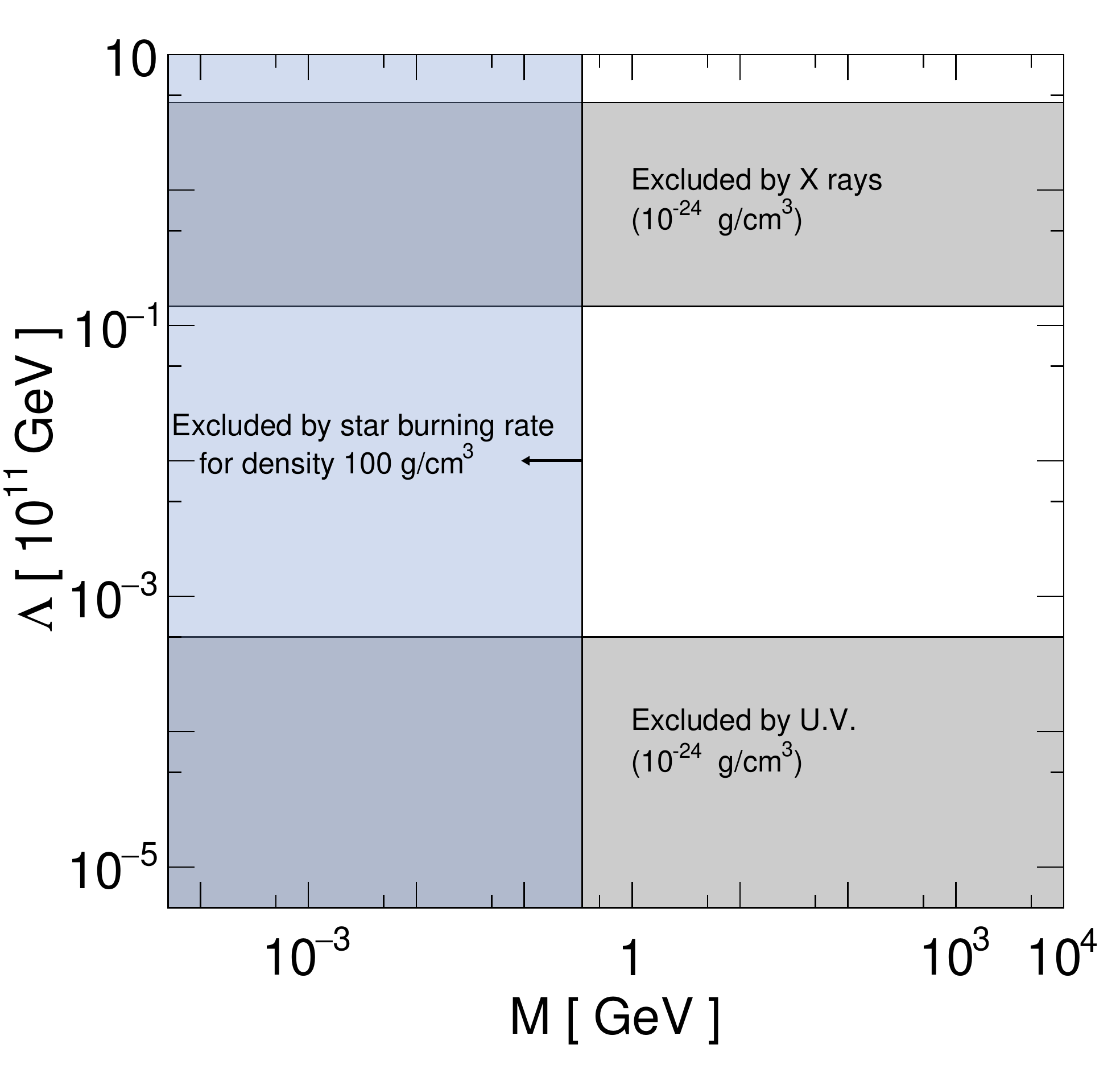}
\caption{Constraint on $\Lambda$ and $M$ at the 95\% C.L. for two ranges of $\Lambda$ values. The grey area shows the constraint which is valid for all masses lower than $4\times 10^{-12}\rm\,eV$.}
\label{fig:constraint_L_M}
\end{figure}

The above-described observations are used to set contraints on the different parameters of the model. For the X-ray spectrum, the method described in section~\ref{sec:statX}, based on Eq.~\ref{eq:maxL}, is used to determine the confidence intervals of the fit and set constraints at the 95\% confidence level (C.L.). For the UV analysis, as stated in Sec.~\ref{sec:statUV}, a $\chi^2$ analysis is performed. All constraints shown in this section are valid for arbitrarily low masses below $4\times10^{-12}\;\rm eV$. In this case, the critical energy $E_\downarrow$ does not depend on the mass of the field and the constraint is therefore independent of this mass. The exclusion regions can be drawn in a $\Lambda-M$ plane to display simultaneously the constraints on the conformal coupling scale and the disformal coupling scale. This is done in Fig.~\ref{fig:constraint_L_M} where the left panel shows a focus on low values of $M$ below $5\times 10^{-5}$ GeV  and the right panel focuses on higher values. In these plots, the grey regions are excluded by the present analysis, and are valid for a matter density of $10^{-24}\; \rm g/cm^3$. As mentioned before, contraints from the star burning rate allow only $M>346$ MeV. However these values being obtained for matter densities of the order of 100 $\rm g/cm^3$, we derive constraints for lower $M$ values, which might still be relevant in the case of a lower matter density. The region between $\Lambda=10^{8}\;\rm GeV$ and  $\Lambda=10^{10}\;\rm GeV$ cannot be probed as it corresponds to the gap between X-ray and UV observations. When $\Lambda$ is set to lower values, only the value of $E_{\downarrow}$ lowers so that spectral irregularities are still present. These orders of magnitudes for the exclusions were anticipated in Fig.~\ref{fig:3} and Fig.~\ref{fig:4}.

\begin{figure}[h]
\includegraphics[width = 0.5\columnwidth]{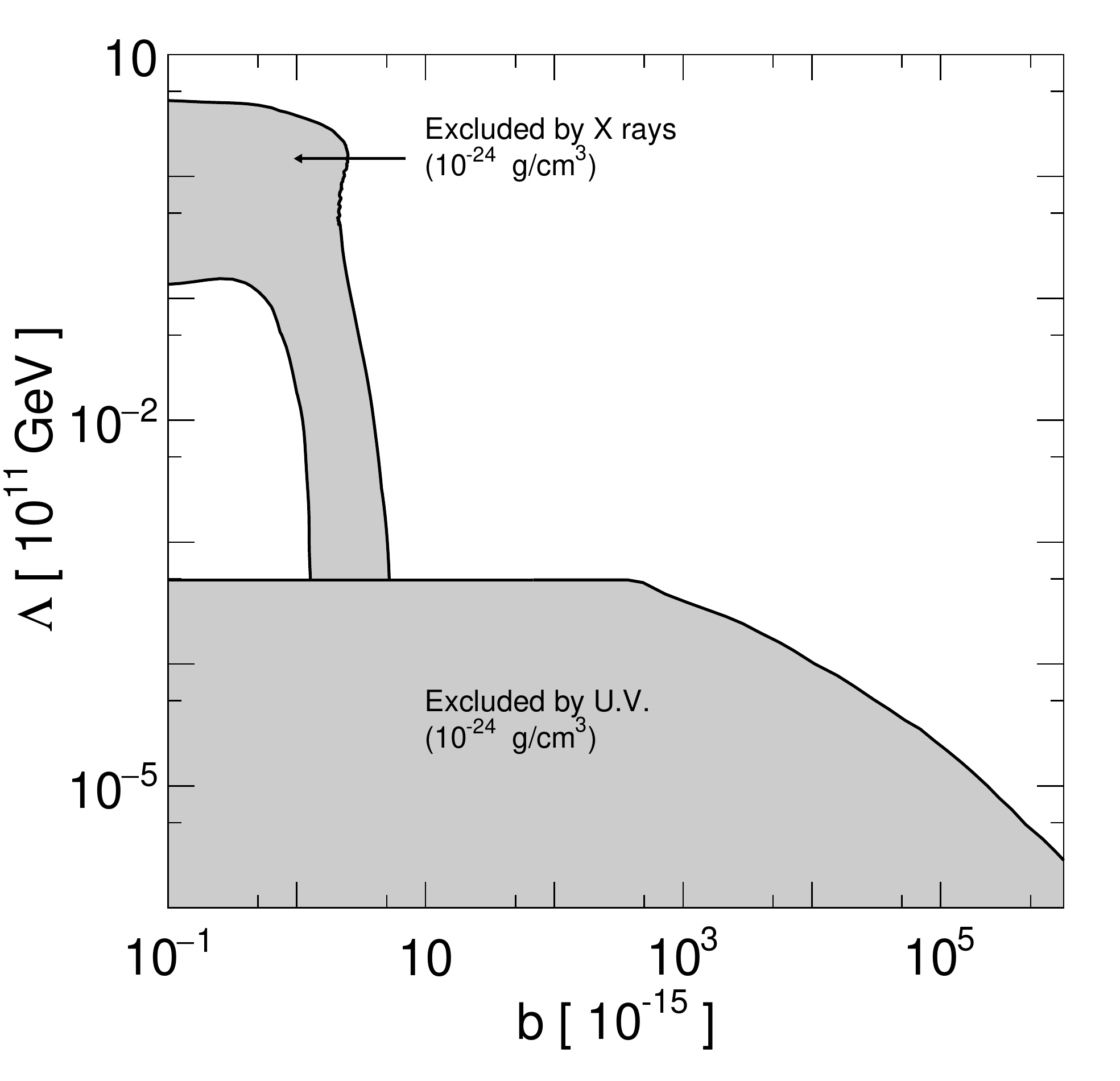}
\caption{Constraint on $\Lambda$ and $b$ at the 95\% C.L. The grey area shows the constraint which is valid for all masses lower than $4\times 10^{-12}\rm\,eV$.}
\label{fig:constraint_L_b}
\end{figure}

Figure~\ref{fig:constraint_L_b} shows the obtained constraints in the $\Lambda$ vs. $b$ plane. Again, the grey areas are excluded, and are valid for masse $m<4\times 10^{-12}\;\rm eV$ and densities of $10^{-24}\;\rm g/cm^3$. In this case, the sensitivity in $\Lambda$ at low $b$ is limited by the value taken by $E_\downarrow$ and not by the statistical uncertainties on the spectral points, as the conversion always occurs in the efficient mixing regime in which the oscillation length is  smaller than the magnetic domain size. When $\Lambda$ goes to higher values, $E_\downarrow$ is shifted to higher energies and no conversion occurs in the energy range of interest. For $b \sim 10^{-12}$, $E_\uparrow$ and $E_\downarrow$ are of the same order of magnitude. The sensitivity in $\Lambda$ decreases roughly as $b^2$ so that $E_\uparrow$ is kept constant, otherwise it is shifted to lower values for higher $b$ and then no conversion occurs. For $b < 10^{-12}$, $E_\uparrow \gg E_\downarrow$ and the ALP case is retrieved and the corresponding constraints on $\Lambda$ hold.

\begin{figure}[h]
\includegraphics[width = 0.5\columnwidth]{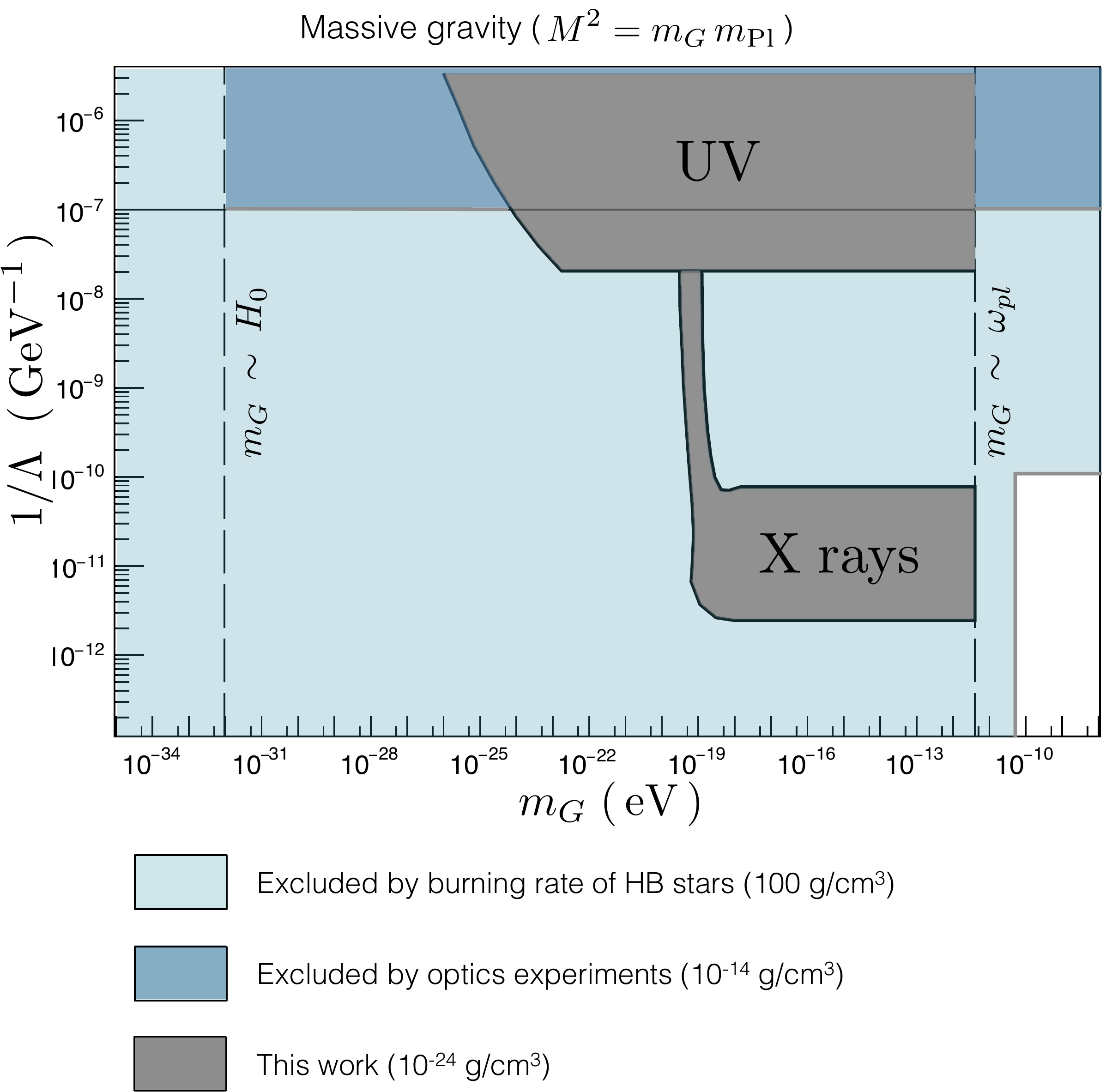}
\caption{Constraint on $1/\Lambda$ and $m$ at the 95\% C.L. for the modified gravity case. Exclusions obtained from the present analysis are shown together with those obtained in different matter density ranges.}
\label{fig:modgrav}
\end{figure}

We have summarised all these constraints in Fig.~\ref{fig:modgrav} where we have chosen the disformal scale $M=\sqrt{m_{\rm Pl} m_G}$, which corresponds to models of massive gravity with a graviton of mass $m_G$. We have represented the results in the $(\Lambda^{-1},m_G)$ plane mimicking the corresponding diagram for ALP's where usually the coupling is the displayed quantity.

First of all, very low masses $m\lesssim H_0$ are the only ones allowed when $\Lambda^{-1}\gtrsim 10^{-7}\ \ {\rm GeV^{-1}}$. The UV exclusion zone that we have obtained in this paper overlaps with the laboratory results from PVLAS and light-shining-through-a-wall experiments (labelled {\it optical experiments} in Fig.~\ref{fig:modgrav}). For lower values of $\Lambda^{-1}$, our X-ray analysis excludes a zone in masses which overlaps with  the  possible masses obtained from the star burning rate and the disformal coupling of scalars to photons, and  where the effect of the direct coupling of scalars to photons implies that $\Lambda^{-1}\le 10^{-10}\ {\rm GeV^{-1}}$. The X-ray constraint is slightly more stringent than the star burning rate bound on $\Lambda^{-1}$. All our new results  from UV and X-ray spectra are only valid for masses below the plasma frequency $\omega_{pl}\sim 4 \times 10^{-12}$ eV. We have also recalled on the same figure the fact that all these bounds have been obtained in different environments and they could be relaxed in models where the disformal coupling becomes density dependent.

\section{Conclusion}

In this study we have considered astrophysical effects of the coupling to radiation of low-mass scalar fields, in a way somewhat comparable to the study of axion-like particle interactions with photons. We have opened up the parameters space to include a disformal coupling, as it would appear in massive gravity models or scalar field-based alternatives to dark energy. Considering photons travelling through turbulent magnetic fields in galaxy clusters, we showed that these interactions could lead to chaotic absorption patterns in the spectra. Such features have been searched for in both UV and X ray bands, with null results. This allowed us to obtain constraints on the coupling scales $\Lambda$ and $M$ of the scalar to radiation, i.e. conformal and disformal couplings respectively. For $\Lambda$ our constraints are complementary to laboratory experiments as we exclude regions between $10^7$ GeV and $10^{11}$ GeV, where dedicated setups are sensitive to scales up to $10^7$ GeV. For $M$, our exclusion limits extend to arbitrarily high energies for certain values of $\Lambda$. They complement both LHC constraints (those are limited to couplings to baryons) and constraints from star burning rates (obtained in a very different range for matter densities). Our limits concern scalar fields with masses below $4\times 10^{-12}\;\rm eV$. If $M$ is interpreted in terms of massive gravity, our analysis allows testing graviton masses down to $m\gtrsim 10^{-23}$ GeV. This means we probe modifications of gravity on scales smaller than one parsec, which are not allowed as long as $\Lambda \lesssim 10^{11}$ GeV. These results partly confirm the exclusions obtained from the study of star burning rates in a totally different matter density range. For larger values of $\Lambda$, X-ray observations are not sufficient and one may envisage gamma-ray probes in order to further constrain the parameter space of photons coupled to light scalar fields.

\acknowledgements
We would like to thank Clare Burrage for her comments on the manuscript.
Part of this work was supported by the French national program PNHE and the ANR project CosmoTeV.
Philippe Brax
acknowledges partial support from the European Union FP7 ITN
INVISIBLES (Marie Curie Actions, PITN- GA-2011- 289442) and from the Agence Nationale de la Recherche under contract ANR 2010
BLANC 0413 01.

\bibliography{bbw}   % name your BibTeX data base

\end{document}